\pdfoutput=1
\documentclass[11pt,twoside,a4paper,cmspaper,final,collab]{cms-tdr}
\def\svnVersion{ae67e3a}\def\svnDate{2024/07/17}\def\cmsCernNoTag{CERN-EP-2023-011}\def\cmsCernDate{\today}\def\cmsMessage{Published in Physical Review Letters as \href{http://dx.doi.org/10.1103/PhysRevLett.133.022302}{\doi{10.1103/PhysRevLett.133.022302}.}}
\begin{document}\cmsNoteHeader{HIN-21-007}

\newcommand{\pp}{\ensuremath{\Pp\Pp}\xspace}
\newcommand{\PbPb}{\ensuremath{\text{PbPb}}\xspace}
\newcommand{\AAa}{\ensuremath{\text{AA}}\xspace}

\newcommand{\RAA}{\ensuremath{R_\AAa}\xspace}
\newcommand{\TAA}{\ensuremath{T_{\AAa}}\xspace}
\newcommand{\NMB}{\ensuremath{N_{\mathrm{MB}}}\xspace}
\newcommand{\Npart}{\ensuremath{N_{\text{part}}}\xspace}
\newlength\cmsTabSkip\setlength{\cmsTabSkip}{1ex}

\ifthenelse{\boolean{cms@external}}{\providecommand{\cmsLeft}{upper\xspace}}{\providecommand{\cmsLeft}{left\xspace}}
\ifthenelse{\boolean{cms@external}}{\providecommand{\cmsRight}{lower\xspace}}{\providecommand{\cmsRight}{right\xspace}}
\ifthenelse{\boolean{cms@external}}{\providecommand{\suppMaterial}
{the supplemental material~\cite{supp}}}
{\providecommand{\suppMaterial}{Appendix~\ref{app:suppMat}}}

\cmsNoteHeader{HIN-21-007}
\title{Observation of the 
\texorpdfstring{\PgUc}{Upsilon(3S)} meson and suppression of \texorpdfstring{\PgU}{Upsilon} states in 
\texorpdfstring{\PbPb}{PbPb} collisions at \texorpdfstring{$\sqrtsNN = 5.02\TeV$}{sqrt(s[NN])=5.02 TeV}}

\author*[cern]{The CMS Collaboration}

\date{\today}

\abstract{The production of \PgUb and \PgUc mesons in lead-lead (\PbPb) and proton-proton (\pp) collisions is studied in their dimuon decay channel using the CMS detector at the LHC. The \PgUc meson is observed for the first time in \PbPb collisions, with a significance above five standard deviations. The ratios of yields measured in \PbPb and \pp collisions are reported for both the \PgUb and \PgUc mesons, as functions of transverse momentum and \PbPb collision centrality. These ratios, when appropriately scaled, are significantly less than unity, indicating a suppression of \PgU yields in \PbPb collisions. This suppression increases from peripheral to central \PbPb collisions. Furthermore, the suppression is stronger for \PgUc mesons compared to \PgUb mesons, extending the pattern of sequential suppression of quarkonium states in nuclear collisions previously seen for the \PJGy, \Pgy, \PgUa, and \PgUb mesons.}

\hypersetup{
pdfauthor={CMS Collaboration},
pdftitle={Observation of the Upsilon(3S) meson and suppression of Upsilon states in PbPb collisions at sqrt(s[NN]) = 5.02 TeV},
pdfsubject={CMS},
pdfkeywords={CMS, heavy ions, Upsilon(3S) observation, Upsilon sequential suppression, QGP}} 

\maketitle 

High-energy heavy ion collisions are useful to study the properties of the quark-gluon plasma (QGP), a strongly coupled medium of deconfined quarks and gluons.
It has long been proposed that the yields of quarkonium states are suppressed because of interactions in the QGP~\cite{Matsui:1986dk,Karsch1988,Kharzeev1994,laine:2007,Brambilla:2008cx, Brambilla:2010vq, blaizot}.
These in-medium effects have been theoretically studied with calculations based on lattice quantum chromodynamics and effective field theories~\cite{Brambilla:2021wkt,Xiaojun:2020,Du:2017qkv,Rothkopf:2020alr}.
One of the fundamental aspects of these interactions with the medium is that the amount of suppression for different quarkonium states is expected to be stronger for those with smaller binding energies.
On the other hand, quarkonia can also be produced by recombination processes~\cite{Thews2001recom, PBMJohanna2000, Brambilla:2021wkt, Brambilla:2016wgg, Emerick:2011xu, Du:2017qkv,blaizot,Xiaojun:2020}.
Studies of \PgU production are particuarly interesting because the number of quark-antiquark pairs in a single lead-lead (\PbPb) collision is much smaller for bottom than for charm quarks, so that the recombination of independently produced quarks and antiquarks can be neglected~\cite{HeavyFlavorQuarkoLHC}.

Experimentally, the dynamics of quarkonium production in heavy ion collisions are commonly studied using the nuclear modification factor (\RAA), which is defined as the ratio of particle yields in nucleus-nucleus (\AAa) collisions to those in proton-proton (\pp) collisions scaled by the average number of binary nucleon-nucleon (NN) interactions in the AA events.

Measurements of \RAA have been performed at the SPS, RHIC, and LHC accelerators, both in the charmonium and bottomonium families~\cite{SPS:Jpsi1997Supp,NA38:1998udo,SPS:Jpsi2000Supp,NA60:2007ewx,PHENIX:2011JpsiAuAu,STAR:2019JpsiAuAu,ALICE:2019JpsiRAA,ALICE:2022psi2sRAA,ATLAS:2018JpsiRAA,HIN16025,HIN16023,ALICE:2021UpsForward,ATLAS:2022UpsRAA,Adamczyk:2013poh,STAR:UpsRAA2022,PHENIX:2015UpsRAA}.
For \PgU states, \RAA results have been reported by CMS~\cite{HIN16023}, ALICE~\cite{ALICE:2021UpsForward}, and ATLAS~\cite{ATLAS:2022UpsRAA} at the LHC, and by STAR~\cite{Adamczyk:2013poh,STAR:UpsRAA2022} and PHENIX~\cite{PHENIX:2015UpsRAA} at RHIC. 
These results show a significant suppression of \PgUa mesons in heavy ion collisions, and \PgUb mesons are even more suppressed.
For the \PgUc meson, only upper limits have been reported in \AAa collisions, by CMS~\cite{HIN16023}.

In this Letter, the first observation of the \PgUc meson in \PbPb collisions is reported.
Nuclear modification factors for \PgUb and \PgUc mesons are found using \PbPb and \pp data collected with the CMS detector at a nucleon-nucleon center-of-mass energy of $\sqrtsNN = 5.02\TeV$ in 2018 and 2017, respectively.
The double ratios, obtained by dividing the \PgUc over \PgUb yield ratios in \PbPb by those for \pp collisions, are also reported.
The \PgU states are identified using their decay into two oppositely charged muons.
The results are presented as functions of \PbPb collision centrality, as well as \PgU meson transverse momentum (\pt) in the rapidity range of $|y|<2.4$.
Centrality is related to the overlap of the two lead nuclei and is quantified as the percentage of the total inelastic nucleus-nucleus hadronic cross section, with 0\% representing the largest overlap~\cite{Enterria:2021ar}.
Tabulated results are provided in the HEPData record for this analysis~\cite{hepdata}.

The CMS apparatus~\cite{CMS:2008xjf} is a multipurpose, nearly hermetic detector, designed to trigger on~\cite{CMS:2020cmk,CMS:2016ngn} and identify electrons, muons, photons, and hadrons~\cite{CMS:2015xaf,CMS:2018rym,CMS:2015myp,CMS:2014pgm}. 
A superconducting solenoid of 6\unit{m} internal diameter provides a magnetic field of 3.8\unit{T}.
Within the solenoid volume are the silicon pixel and strip tracker, a crystal electromagnetic calorimeter, and a brass-scintillator hadron calorimeter.
The pseudorapidity ($\eta$) coverage of these calorimeters is extended by the forward hadron (HF) calorimeters, located at $3 < \abs{\eta} < 5$.
Muons are measured in the range $\abs{\eta} < 2.4$ using gas-ionization detectors embedded in the steel flux-return yoke outside the solenoid.
Events are filtered using a two-tiered trigger system~\cite{CMS:2016ngn}. The first level (L1), composed of custom hardware processors, uses information from the calorimeters and muon detectors~\cite{CMS:2020cmk}.
The second level, known as the high-level trigger (HLT), consists of a farm of processors running a version of the full event reconstruction software.
Centrality is determined using the sum of the total transverse energy deposited in both of the HF calorimeters~\cite{Chatrchyan:2011sx}.

The events are selected online using an L1 trigger requiring two muons in a single bunch crossing without explicit requirements on the muon momentum.
Additional criteria on the single-muon quality and dimuon mass selection are applied at the HLT in \PbPb collisions~\cite{HIN19002}.
The collected \pp (\PbPb) sample corresponds to an integrated luminosity of 300\pbinv (1.61\nbinv)~\cite{CMS-PAS-LUM-19-001,CMS-PAS-LUM-18-001,CMS:2021LumEPJC}.

In order to reject beam-gas interactions and nonhadronic collisions, an offline event selection is applied for \pp and \PbPb collisions~\cite{CMS:2017JHEP_ChargedParticleRAA}. For both systems, events are required to have at least one reconstructed primary vertex (PV), which is the vertex corresponding to the hardest scattering in the event, evaluated using tracking information alone~\cite{CMS-TDR-15-02}.
In addition, more than 25\% of the tracks have to pass a tight track-quality selection in \pp collisions~\cite{CMS:2017JHEP_ChargedParticleRAA,CMS:2014pgm}. For \PbPb collisions, selections are applied on the silicon pixel detector cluster widths~\cite{CMS:2017JHEP_ChargedParticleRAA,CMS:2010za} and on the HF information~\cite{HIN19002}.
The reconstructed muons are selected using criteria that optimize the muon identification~\cite{HIN19002}. In addition, to ensure high reconstruction efficiency for \PgU mesons, the individual muons are required to have $\pt>3.5$\GeVc and $|\eta|<2.4$.

Monte Carlo (MC) samples for each \PgU meson are simulated using the ``CP5" tune~\cite{cp5tune} of \PYTHIA 8.212~\cite{Sjostrand:2014zea}, with the assumption that they are produced unpolarized~\cite{CMS:UpsPolarization,CMS:UpsPolarizationMult,LHCb:UpsPolarization,Acharya:2021PLB}.
To reproduce the background environment in \PbPb collisions, each \PYTHIA \PgU event is embedded into a \PbPb event simulated using \HYDJET 1.9~\cite{Lokhtin:2005px}.
The MC events are then weighted to match the measured \PgU \pt spectra for either \pp or \PbPb collisions~\cite{HIN19002}.
A full simulation of the CMS detector using \GEANTfour~\cite{Agostinelli:2002hh} is performed to determine the acceptance and reconstruction efficiency. 
The feed-down contributions, i.e., decays from heavier quarkonium states, are not considered in this analysis.
The effect of such contributions on the kinematic distributions of the simulated \PgU states is, to a large extent, accounted for by the weighting procedure. 

The dimuon invariant mass distribution is studied in the 8--14\GeVcc region.
To reduce the large amount of background in \PbPb collisions, signal-enriched dimuon candidates are selected using boosted decision trees (BDTs), using the \textsc{tmva} package~\cite{Hocker:2007ht}.
For the BDT training, MC samples and dimuons from the invariant mass spectra in data except where the \PgU signals are present (i.e., 8.8--10.8\GeVcc) are used for the signal and background, respectively.
In both cases, the dimuons must satisfy the selection criteria previously mentioned.
The training variables for the BDT include:
the $\chi^2$ probability of the dimuon vertex fit; 
the distance of closest approach of the \PgU meson momentum vector relative to the PV;
the distance between the PV and the dimuon vertex, both in three dimensions and projected onto the transverse plane;
the pointing angle, defined as the angle between the line segment connecting the PV and decay vertex and the momentum vector of the reconstructed particle candidates, again in three dimensions and projected onto the transverse plane.
To avoid potential biases, the algorithms and selections are optimized using a quarter of the \PbPb data for dimuon $\pt < 30\GeVc$, $\abs{y}<2.4$, and centrality 0--90\%.
A threshold is set on the resulting BDT variable to maximize the signal significance, $\mathrm{S} / \sqrt{ \mathrm{ S } + \mathrm{ B } }$, where $\mathrm{S}$ and $\mathrm{B}$ represent the yields for signal and background, respectively. 

The yields of the \PgU states are extracted by an extended unbinned maximum likelihood fit to the dimuon invariant mass distributions.
The line shape of each \PgU state is modeled by a sum of three Crystal Ball (CB) functions~\cite{SLAC-R-236}, with common means and tail parameters but independent widths.
The means and widths for the \PgUb and \PgUc are found by multiplying the fitted values of those for the \PgUa by the ratio of their world-average masses~\cite{PDG2022} over that for the \PgUa.
All other fit parameters, except the yields for the excited states, are fixed to those found for the ground state.
The shape parameters of the signal fit model, with the exception of the \PgUa mean, are constrained by Gaussian probability density functions (pdfs), of means and widths determined, for each \pt bin, from the central values and uncertainties obtained from fits to the MC samples.
The background of the mass distribution is modeled using three different functional forms: an error function multiplied by an exponential, a simple exponential, and a Chebyshev polynomial.
An Akaike information criterion (AIC) test~\cite{AIC:1100705}, which estimates the relative quality of each function, is performed to determine, for each kinematic region, the nominal function for the background pdf.
The order of the Chebyshev polynomial is chosen based on a log likelihood ratio test~\cite{LLR:3277}.

\begin{figure}[htb]
 \centering
   \includegraphics[width=0.49\textwidth]{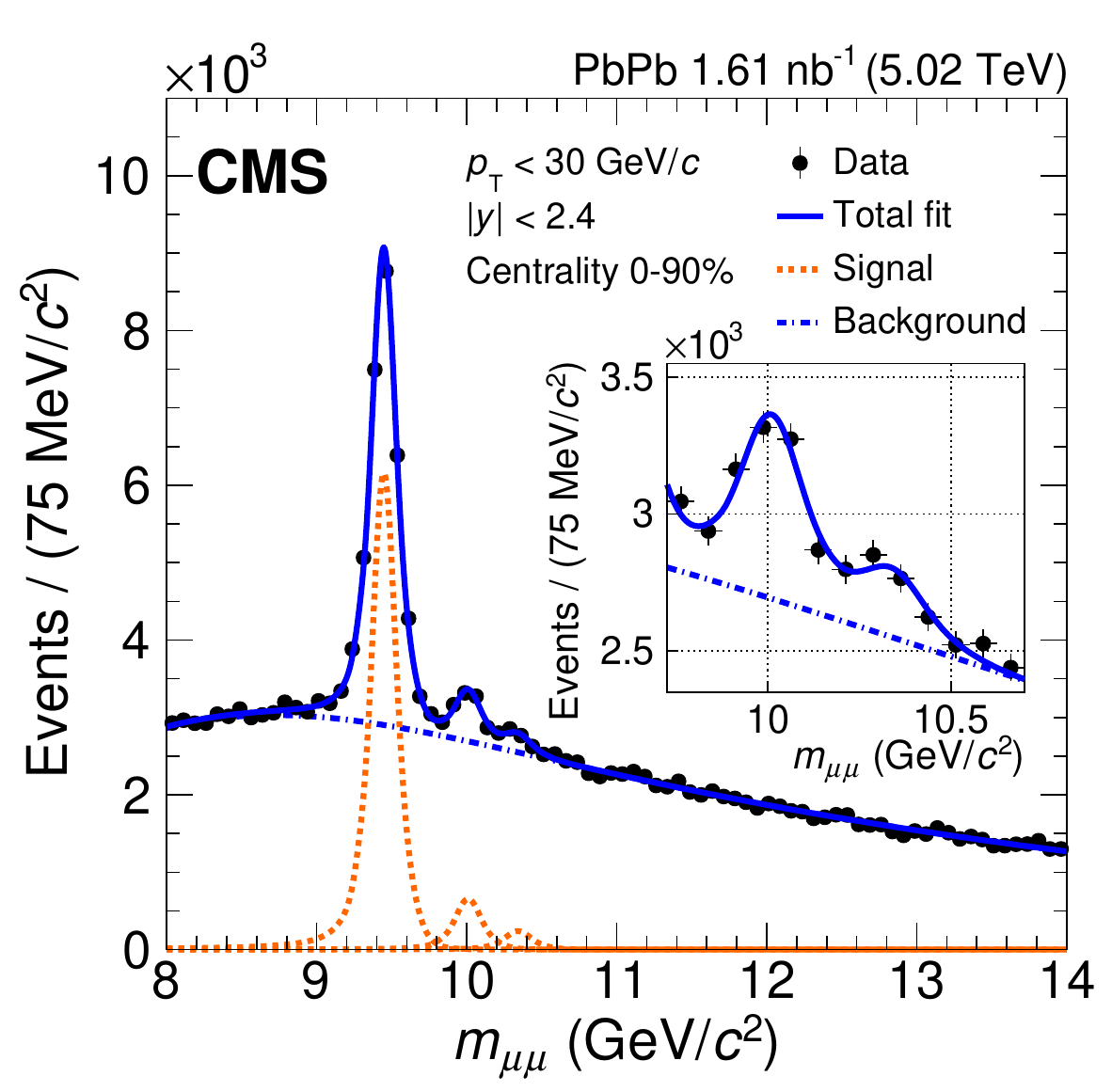}
   \caption{Dimuon invariant mass distribution in \PbPb collisions, integrated over the full kinematic range $\pt < 30$\GeVc and $\abs{y}<2.4$. The solid curves show the result of the fit, whereas the orange dashed and blue dash-dotted curves represent the three \PgU states and the background, respectively. The inset shows the region around the mass of the \PgUc meson.}
  \label{fig:sigFit}
\end{figure}

Figure~\ref{fig:sigFit} shows the dimuon invariant mass distribution in \PbPb collisions, integrated over the full kinematic range, $\pt < 30$\GeVc and $\abs{y}<2.4$, and centrality 0--90\%. 

Acceptance and efficiency correction factors are applied to the extracted number of \PgU mesons to compute the \RAA values.
The acceptance is computed with simulated samples, as the fraction of generated \PgU mesons that decay to muons of $\pt^{\mu} > 3.5\GeVc$ and $\abs{\eta^{\mu}} < 2.4$.
The efficiency is also evaluated with simulated events, as the fraction of accepted dimuons that are reconstructed and pass the trigger and analysis selection criteria.
To take into account possible differences between data and MC simulations, the individual components of the dimuon efficiency (reconstruction, identification, and triggering) are measured using single muons from \PJGy meson decays in both data and simulation, with the \textit{tag-and-probe} (T\&P) method~\cite{CMS:2021jinst}.
The data over simulation ratios, for each of the three components, are applied as event-by-event weights to each dimuon. The acceptance and efficiency values for the full kinematic region are 40\% (43\%) and 37\% (38\%) for \PgUb (\PgUc) in \PbPb collisions.

The systematic uncertainties are analyzed for various sources and summarized in Table~\ref{tab:syst}. For each source, the difference of the signal yields in the variation compared to the nominal is taken as the systematic uncertainty.
Three sources are considered for the uncertainty in the signal extraction: choice of signal pdf parameters, choice of signal pdf, and choice of background pdf.
For the parameterization of the signal pdf, the fit results using either the \PgUb or \PgUc MC samples are used to determine different values for the mean and width of the Gaussian function for each signal parameter.
The signal pdf is modified from a sum of three CB functions to a sum of two CB functions and a Gaussian function.
For the background pdf systematic study, functions rejected by the AIC test in the nominal background pdf determination, for each kinematic region, are used as alternatives.

The acceptance and efficiency uncertainties are evaluated by varying the \pt weights (used to match the simulated and measured \PgU \pt spectra) by their fit uncertainties.
The uncertainty on the T\&P correction of the single muon efficiencies is propagated to the dimuon efficiencies.

For the \PbPb analysis, additional uncertainties arise from the BDT training and centrality calibration.
To evaluate the uncertainty reflecting the BDT training, the event samples used as signal or background are split in two.
Each trained algorithm is applied to the other half of the samples and the average of the two BDT variable values is used as the nominal.
Alternatively, the two individual values of each BDT variable are used and the largest difference of the corrected signal yields compared to the nominal is taken as the systematic uncertainty.
The centrality calibration is varied by changing the boundaries of the centrality intervals~\cite{HIN19002} considering the inefficiency in the event selection.

The total uncertainty, dominated by the uncertainties from the BDT training and background pdf, is the quadratic sum of the uncertainties from all the different sources. 
In addition, global uncertainties reflect the integrated luminosity of the \pp data set (1.9\%)~\cite{CMS-PAS-LUM-19-001} and the number of minimum bias \PbPb collision events (\NMB) sampled by the trigger (1.3\%) ~\cite{HIN19007}.
The overlap function \TAA for each centrality interval is the average number of binary NN collisions per \PbPb interaction divided by the inelastic NN cross section. Its uncertainty is estimated by varying the Glauber model parameters within their uncertainties~\cite{PhysRevC.97.054910} and is found to be in the range of 1.8--5.4\%.

\begin{table*}[htb]
  \centering
  \topcaption{Systematic uncertainties from various sources in $\Pp\Pp$ and \PbPb collisions listed in percentage. The global uncertainties described in the text are not included in the total uncertainties.}
  \begin{scotch}{lcccccc}
  & \multicolumn{2}{c}{\PgUb(\%)}                                   & \multicolumn{2}{c}{\PgUc(\%)}                                          & \multicolumn{2}{c}{\PgUc/\PgUb(\%)}                             \\ 
\multicolumn{1}{l}{Source}             & \multicolumn{1}{c}{pp}             & \multicolumn{1}{c}{PbPb}                    	  & \multicolumn{1}{c}{pp}  	    & \multicolumn{1}{c}{PbPb}           & \multicolumn{1}{c}{pp}                     & \multicolumn{1}{c}{PbPb} \\ \hline
\multicolumn{1}{l}{BDT selection}      & \multicolumn{1}{c}{\NA}              & \multicolumn{1}{c}{0.3--9.0} 				   	& \multicolumn{1}{c}{\NA}              & \multicolumn{1}{c}{1.5--18.6}        & \multicolumn{1}{c}{\NA}        & \multicolumn{1}{c}{1.2--22.8} \\ 
\multicolumn{1}{l}{Background PDF}     & \multicolumn{1}{c}{0.1--1.4}        & \multicolumn{1}{c}{0.3--11.7}				   	 & \multicolumn{1}{c}{0.2--1.6}        & \multicolumn{1}{c}{1.4--21.4}     & \multicolumn{1}{c}{\textless{}0.5} & \multicolumn{1}{c}{0.6--17.6} \\ 
\multicolumn{1}{l}{Signal PDF}         & \multicolumn{1}{c}{0.1--1.1}        & \multicolumn{1}{c}{0.5--2.6} 				   	& \multicolumn{1}{c}{0.4--1.1}        & \multicolumn{1}{c}{0.1--2.5}        & \multicolumn{1}{c}{0.3--0.6}     & \multicolumn{1}{c}{0.1--3.0} \\ 
\multicolumn{1}{l}{Signal parameter}   & \multicolumn{1}{c}{0.1--1.2}        & \multicolumn{1}{c}{0.0--3.8}				   	 & \multicolumn{1}{c}{0.1--1.6}        & \multicolumn{1}{c}{0.3--3.7}        & \multicolumn{1}{c}{0.05--1.4}     & \multicolumn{1}{c}{0.1--0.9} \\
\multicolumn{1}{l}{Event selection}    & \multicolumn{1}{c}{\NA}              & \multicolumn{1}{c}{0.0--0.5} 				   	& \multicolumn{1}{c}{\NA}              & \multicolumn{1}{c}{0.2--13.1}        & \multicolumn{1}{c}{\NA}           & \multicolumn{1}{c}{0.1--13.6} \\ 
\multicolumn{1}{l}{Correction factors} & \multicolumn{1}{c}{\textless{}0.1} & \multicolumn{1}{c}{\textless{}0.5} 		   	& \multicolumn{1}{c}{\textless{}0.1} & \multicolumn{1}{c}{\textless{}0.4}        & \multicolumn{2}{c}{\textless{}2.0}             \\ 
\multicolumn{1}{l}{T\&P}               & \multicolumn{1}{c}{0.9--1.0}        & \multicolumn{1}{c}{3.8--4.5} 				   	& \multicolumn{1}{c}{0.9--1.1}        & \multicolumn{1}{c}{3.8--4.4}        & \multicolumn{2}{c}{\NA} \\ 
\multicolumn{1}{l}{Total uncertainty}  & \multicolumn{1}{c}{1.0--1.8}        & \multicolumn{1}{c}{3.9--13.5} 				   	& \multicolumn{1}{c}{1.1--2.2}        & \multicolumn{1}{c}{6.0--22.2}        & \multicolumn{1}{c}{0.4--1.5}     & \multicolumn{1}{c}{4.1--23.8} 
\end{scotch}
\label{tab:syst}
\end{table*}

The significance of the \PgUc in \PbPb collisions corresponds to 5.6 standard deviations, calculated using the fit likelihood ratio~\cite{Wilks:1938dza,Cowan:2010js,Disllh2014}.
The nuclear modification factors are computed as
\begin{linenomath}
\begin{equation}
\RAA (\pt) = \frac{ N_{\textrm{AA}}(\pt) }{ \langle \TAA \rangle \sigma^{\pp}(\pt) },
\end{equation}
\end{linenomath}
where $N_{\textrm{AA}}$ and $\sigma^{\Pp\Pp}$ are the efficiency- and acceptance-corrected normalized yields in \PbPb collisions and the \pp cross sections, respectively, for \PgU mesons in a given kinematic range. The average value of \TAA computed in each centrality bin is denoted by $\langle \TAA \rangle$.

\begin{figure}[hbtp]
  \centering
    \includegraphics[width= 0.503\textwidth]{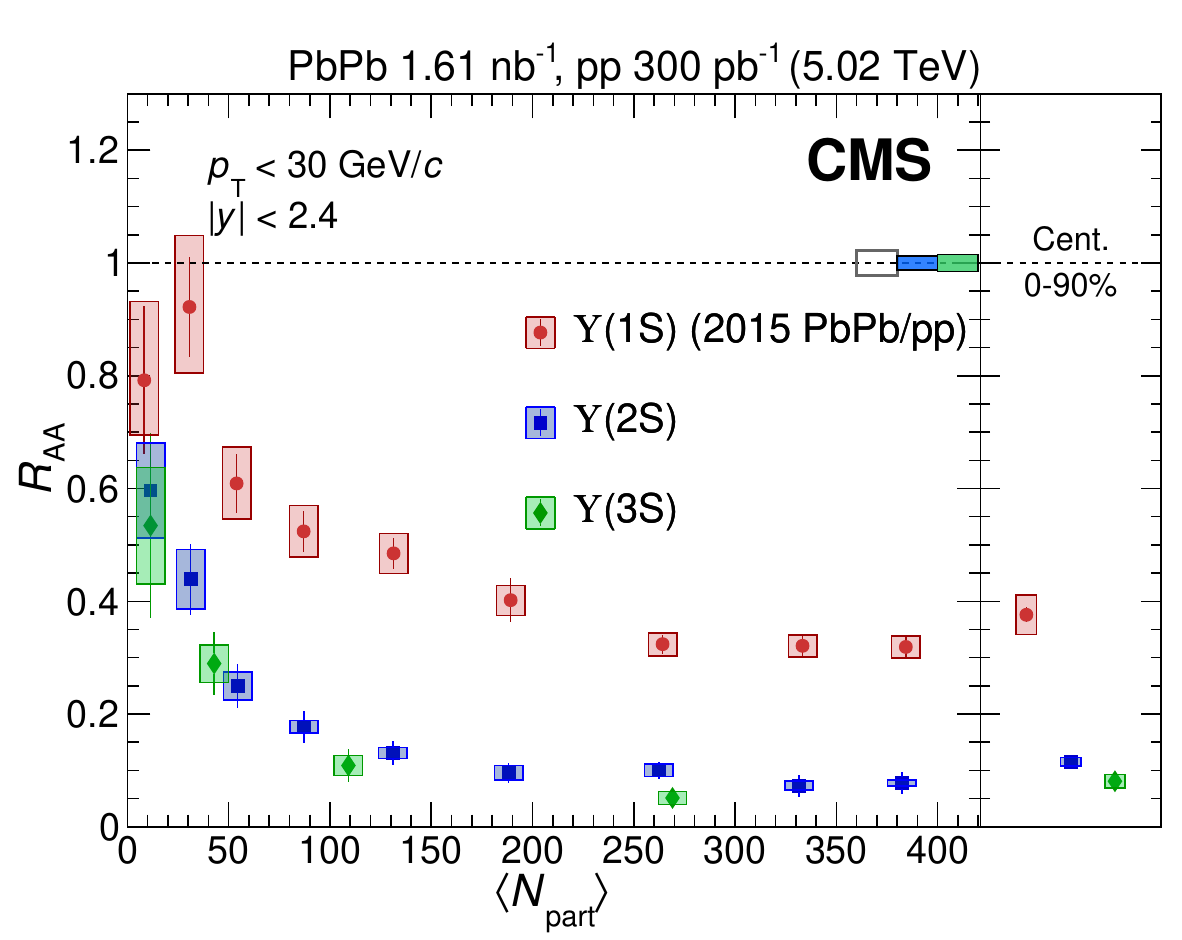}
    \includegraphics[width= 0.457\textwidth]{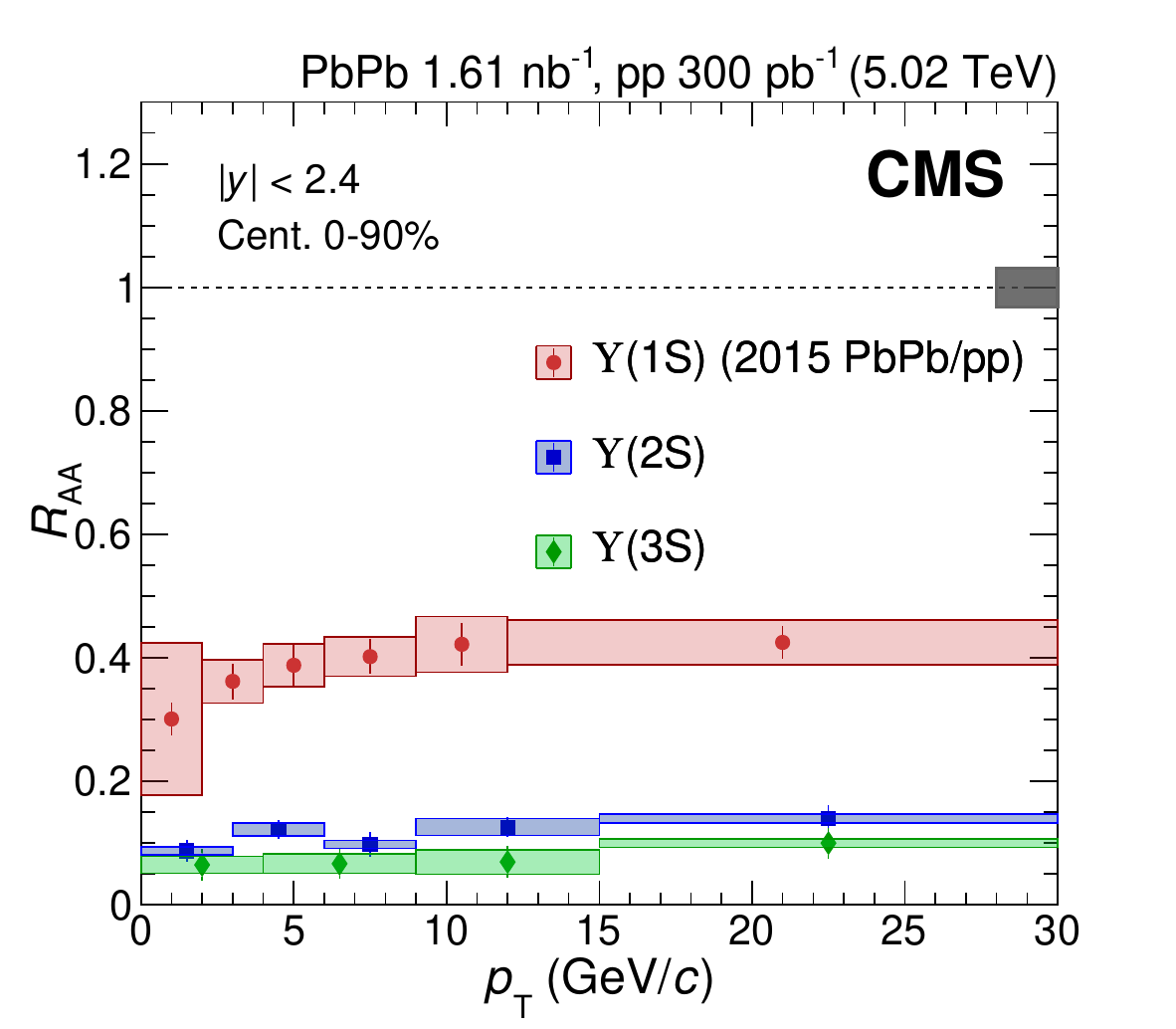}
    \caption{Measured \RAA for the \PgU states as functions of $\langle \Npart \rangle$ (\cmsLeft), showing also the \mbox{0--90\%} centrality interval, and of \pt (\cmsRight). The vertical lines and boxes correspond to statistical and systematic uncertainties, respectively. In the \cmsLeft plot, the leftmost box at unity represents the \pp luminosity and \PbPb \NMB combined uncertainties, whereas the second (third) box corresponds to the uncertainty on the \PgUb (\PgUc) \pp yields. The box at unity in the \cmsRight plot combines the uncertainties of \TAA, \pp luminosity, and \PbPb \NMB. The results for the \PgUa are taken from Ref.~\cite{HIN16023} and are not affected by the boxes at unity.}
    \label{fig:resRAA}
\end{figure}

Figure~\ref{fig:resRAA} presents the \RAA values for \PgUb and \PgUc, together with the previous measurements for the \PgUa~\cite{HIN16023}, as functions of \PgU meson \pt and \PbPb collision centrality, with the latter represented using $\langle \Npart \rangle$, the average number of participating nucleons in each of the centrality intervals~\cite{PhysRevC.97.054910}.
The measurements as a function of \pt are reported in five intervals for the \PgUb, 0--3, 3--6, 6--9, 9--15, and 15--30\GeVc, and four for the \PgUc, 0--4, 4--9, 9--15, and 15--30\GeVc.
The centrality intervals and the corresponding $\langle \Npart \rangle$ values used for this analysis are listed in the HEPData record~\cite{hepdata}.
The results for the \PgUa are taken from Ref.~\cite{HIN16023} because the new data do not improve the statistical significance. Furthermore, the systematic uncertainties, which are dominant for the \PgUa results, are found to not improve when including the BDT training.

A gradual decrease of \RAA is observed towards more central collisions (i.e., higher \Npart) for both \PgUb and \PgUc.
On the other hand, no significant dependence on \pt is found for either of them.
Both states are strongly suppressed in central \PbPb collisions, as well as over the entire \pt region when averaged over centrality. Furthermore, the \RAA of the \PgUc is smaller than that of the \PgUb, with values integrated over \pt and centrality 0--90\% of 0.115 $\pm$ 0.008 (stat) $\pm$ 0.007 (syst) and 0.080 $\pm$ 0.014 (stat) $\pm$ 0.012 (syst) for the \PgUb and \PgUc, respectively. 
These results indicate, much more clearly than the previous measurements~\cite{HIN16023,ALICE:2021UpsForward,ATLAS:2022UpsRAA}, that the sequential suppression pattern of the bottomonium states follows the ordering of their binding energies~\cite{FaccioliLourenco:2018epjc}.

\begin{figure}[hbtp]
	\centering
       		 \includegraphics[width= 0.503\textwidth]{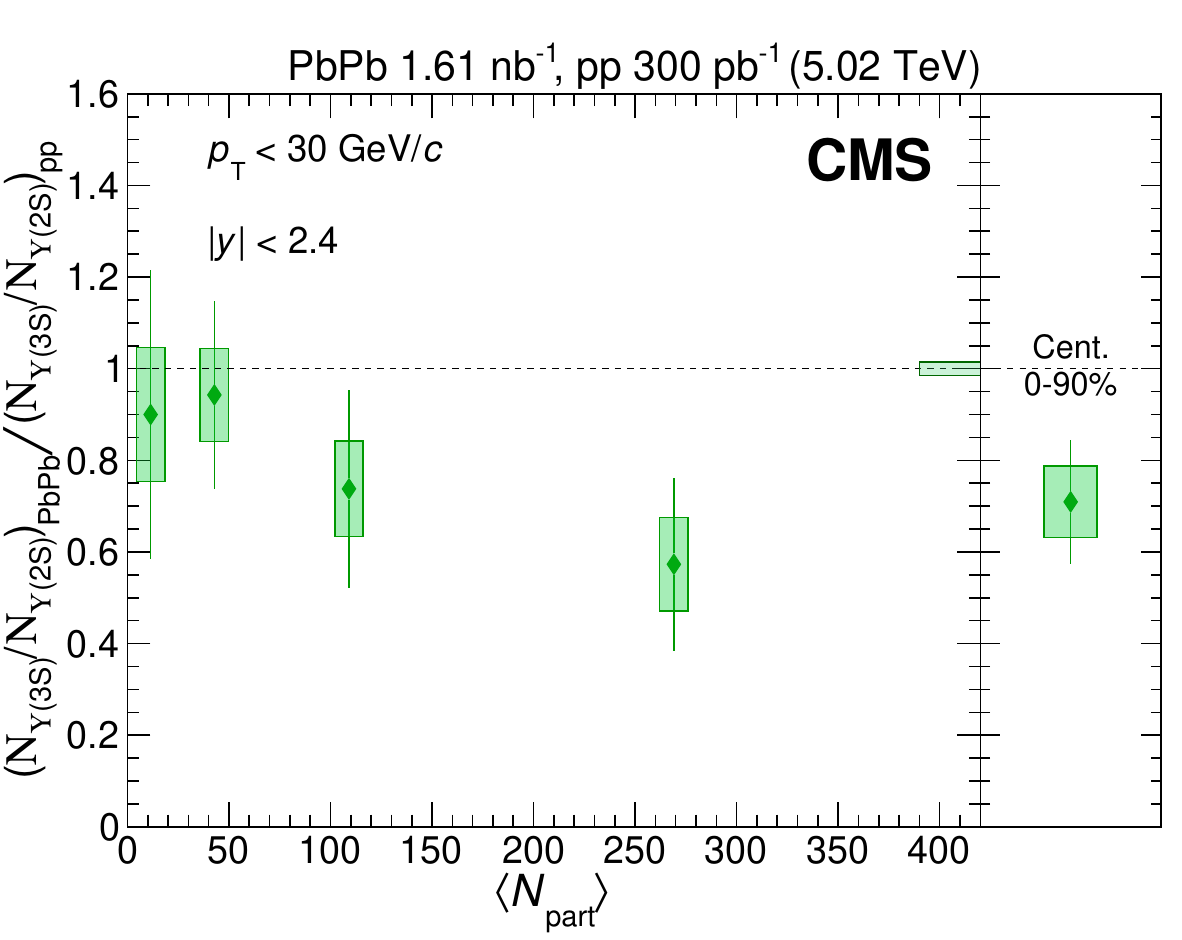}
       		 \includegraphics[width= 0.457\textwidth]{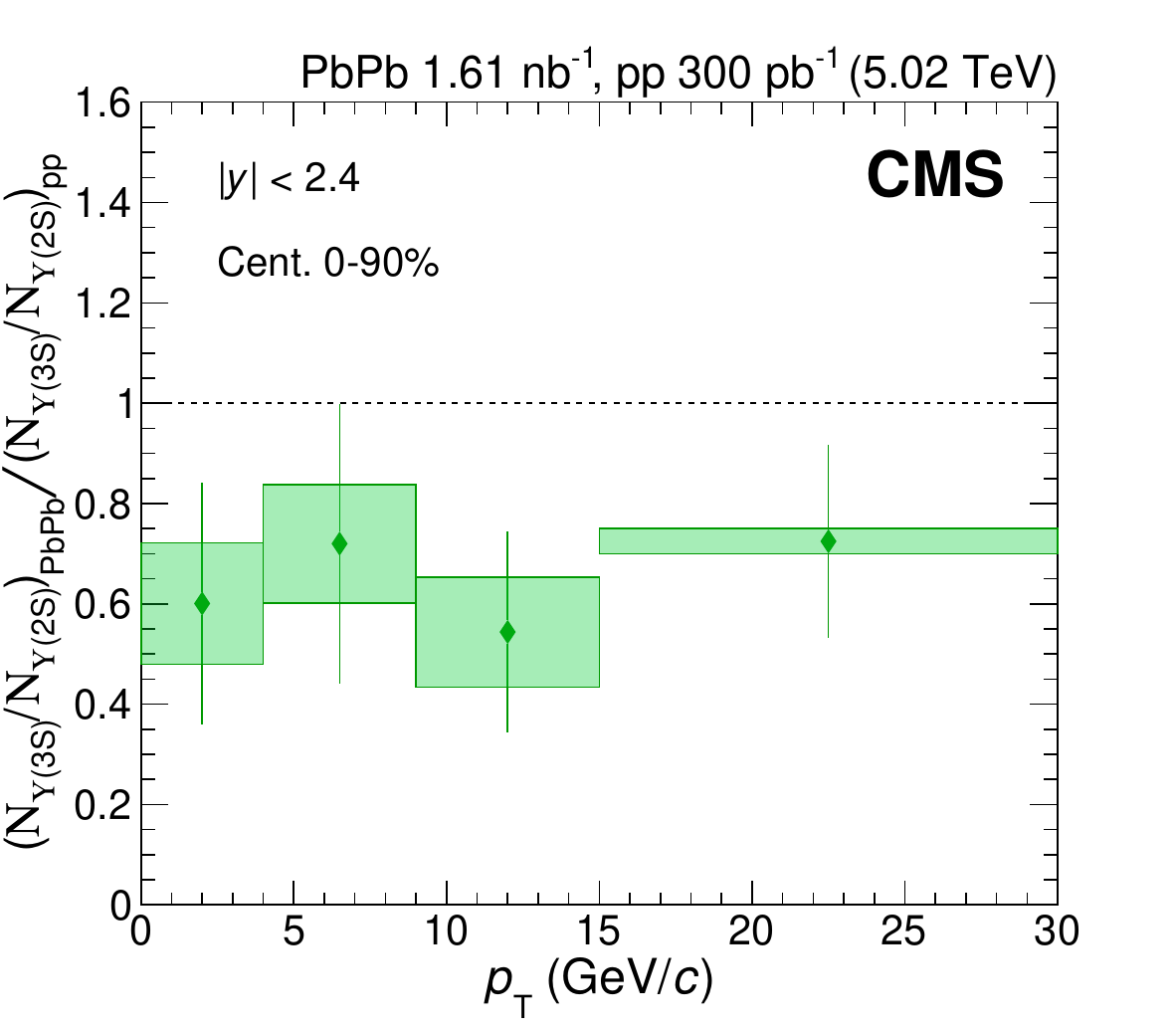}
          \caption{The double ratios of \PgUc/\PgUb as functions of $\langle \Npart \rangle$ (\cmsLeft), showing also the \mbox{0--90\%} centrality interval, and of \pt (\cmsRight). The vertical lines and boxes correspond to statistical and systematic uncertainties, respectively. The box at unity in the \cmsLeft plot shows the combined systematic and statistical uncertainties from \pp data, which is common to all the points.}
        	\label{fig:resDR}
\end{figure}

Figure~\ref{fig:resDR} shows, as functions of $\langle \Npart \rangle$ and \pt, the \PgUc to \PgUb double ratios, obtained by dividing the \PgUc over \PgUb yield ratios in \PbPb collisions by the corresponding ratios in \pp collisions. 
The bins are identical to those used in the \PgUc measurement. 
The uncertainties in the \PgUc/\PgUb ratios are computed by propagating the \PgUb and \PgUc uncertainties, taking into account their correlation, whereas the global uncertainties on \TAA, \pp luminosity, and \PbPb \NMB cancel out.
The decreasing trend of the double ratios towards more central \PbPb collisions indicates a stronger suppression for the \PgUc than for the \PgUb, whereas no significant dependence on \pt is seen. Comparisons to various theoretical calculations ~\cite{Xiaojun:2020,Brambilla:2023Reg,FerreiroComp,Du:2017qkv,Wolschin:2020ijmpa,Andronic2018} and previous experimental results ~\cite{ALICE:JpsiRAA2019,Sirunyan:2018mp,ALICE:2022psi2sRAA,BindingESatz,HIN18005} can be found in \suppMaterial{}.

In summary, data from \PbPb and \pp collisions at a nucleon-nucleon center-of-mass energy of $\sqrtsNN = 5.02\TeV$, collected with the CMS detector, were analyzed to measure the yields and nuclear modification factors, \RAA, of the \PgUb and \PgUc mesons.
The \PgUc meson is observed for the first time in \PbPb collisions, with a significance above five standard deviations.
Dividing the \PgUc over \PgUb yield ratios in \PbPb by those in \pp collisions gives the double ratios that quantify the relative modification of the two mesons.
Results are shown as functions of \PgU transverse momentum and \PbPb collision centrality.
Both the \PgUb and \PgUc mesons are suppressed ($\RAA<1$), with a stronger effect for the \PgUc.
The suppression increases for more central \PbPb collisions, whereas no significant dependence on \pt is seen.
The \PgUc over \PgUb double ratios show a gradual decrease towards more central \PbPb collisions, indicating that the degree to which the suppression is stronger for the \PgUc meson increases for more central \PbPb collisions.
Combined with previous measurements, these results indicate that the strength of the suppression increases in the sequence \PgUa, \PgUb, and \PgUc.
These results provide new constraints on the understanding of the dynamics of quarkonium states in the QGP created in heavy ion collisions. 

\begin{acknowledgments}
  We congratulate our colleagues in the CERN accelerator departments for the excellent performance of the LHC and thank the technical and administrative staffs at CERN and at other CMS institutes for their contributions to the success of the CMS effort. In addition, we gratefully acknowledge the computing centers and personnel of the Worldwide LHC Computing Grid and other centers for delivering so effectively the computing infrastructure essential to our analyses. Finally, we acknowledge the enduring support for the construction and operation of the LHC, the CMS detector, and the supporting computing infrastructure provided by the following funding agencies: BMBWF and FWF (Austria); FNRS and FWO (Belgium); CNPq, CAPES, FAPERJ, FAPERGS, and FAPESP (Brazil); MES and BNSF (Bulgaria); CERN; CAS, MoST, and NSFC (China); MINCIENCIAS (Colombia); MSES and CSF (Croatia); RIF (Cyprus); SENESCYT (Ecuador); MoER, ERC PUT and ERDF (Estonia); Academy of Finland, MEC, and HIP (Finland); CEA and CNRS/IN2P3 (France); BMBF, DFG, and HGF (Germany); GSRI (Greece); NKFIH (Hungary); DAE and DST (India); IPM (Iran); SFI (Ireland); INFN (Italy); MSIP and NRF (Republic of Korea); MES (Latvia); LAS (Lithuania); MOE and UM (Malaysia); BUAP, CINVESTAV, CONACYT, LNS, SEP, and UASLP-FAI (Mexico); MOS (Montenegro); MBIE (New Zealand); PAEC (Pakistan); MES and NSC (Poland); FCT (Portugal); MESTD (Serbia); MCIN/AEI and PCTI (Spain); MOSTR (Sri Lanka); Swiss Funding Agencies (Switzerland); MST (Taipei); MHESI and NSTDA (Thailand); TUBITAK and TENMAK (Turkey); NASU (Ukraine); STFC (United Kingdom); DOE and NSF (USA).
\end{acknowledgments}

\bibliography{auto_generated}
\ifthenelse{\boolean{cms@external}}{}{
\numberwithin{figure}{section}
\numberwithin{table}{section}
\appendix
\section{Supplemental material\label{app:suppMat}}

\begin{figure}[!h]
\centering
  \includegraphics[width=0.48\textwidth]{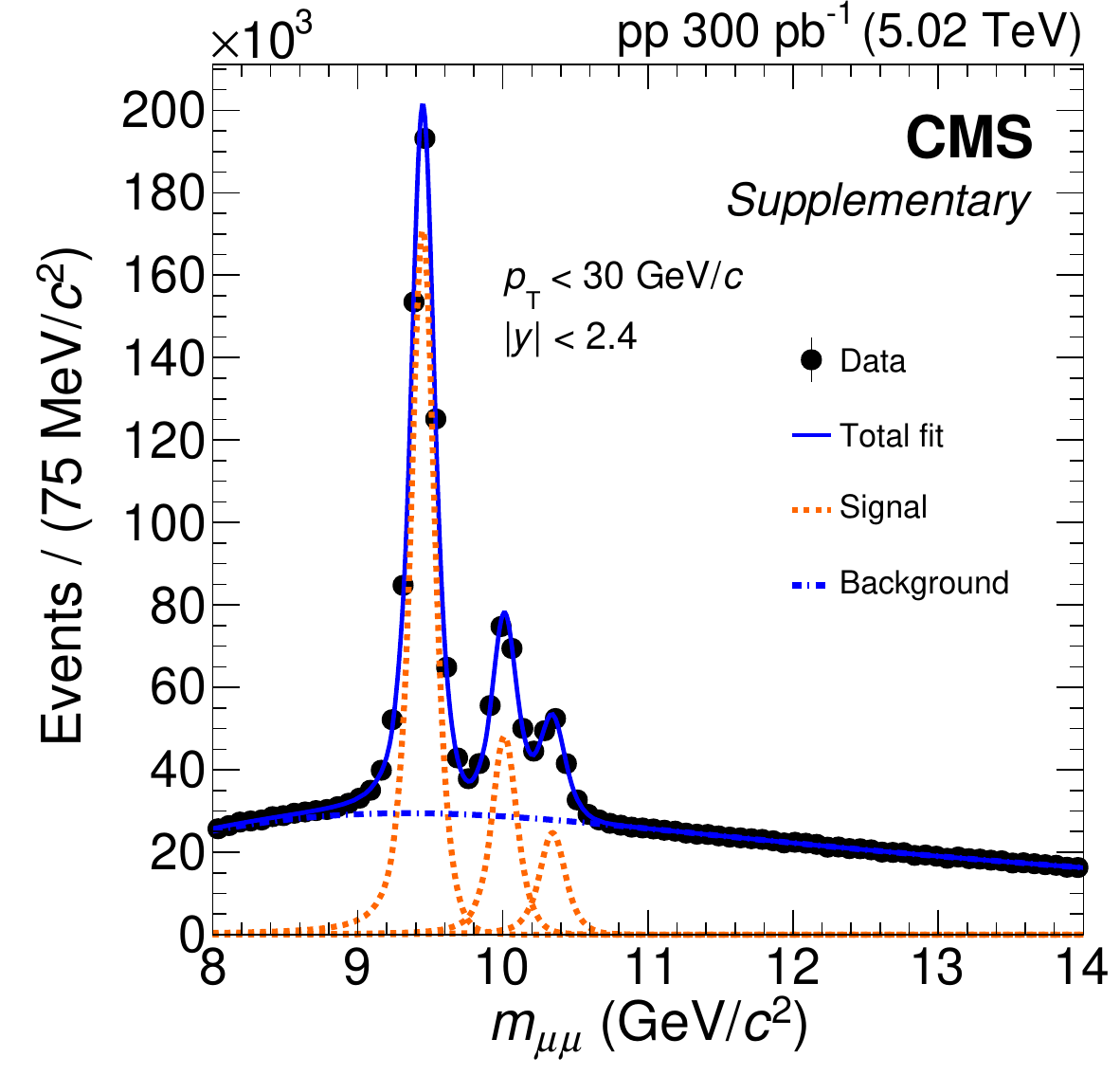}
  \caption{Dimuon invariant mass distribution in \pp collisions, integrated over the full kinematic range $\pt < 30$\GeVc and $\abs{y}<2.4$. The solid curves show the result of the fit, whereas the orange dashed and blue dash-dotted curves represent the three \PgU states and the background, respectively.}
  \label{fig:sigFitpp}
\end{figure}

\begin{figure}[!h]
\centering
\includegraphics[width=0.48\textwidth]{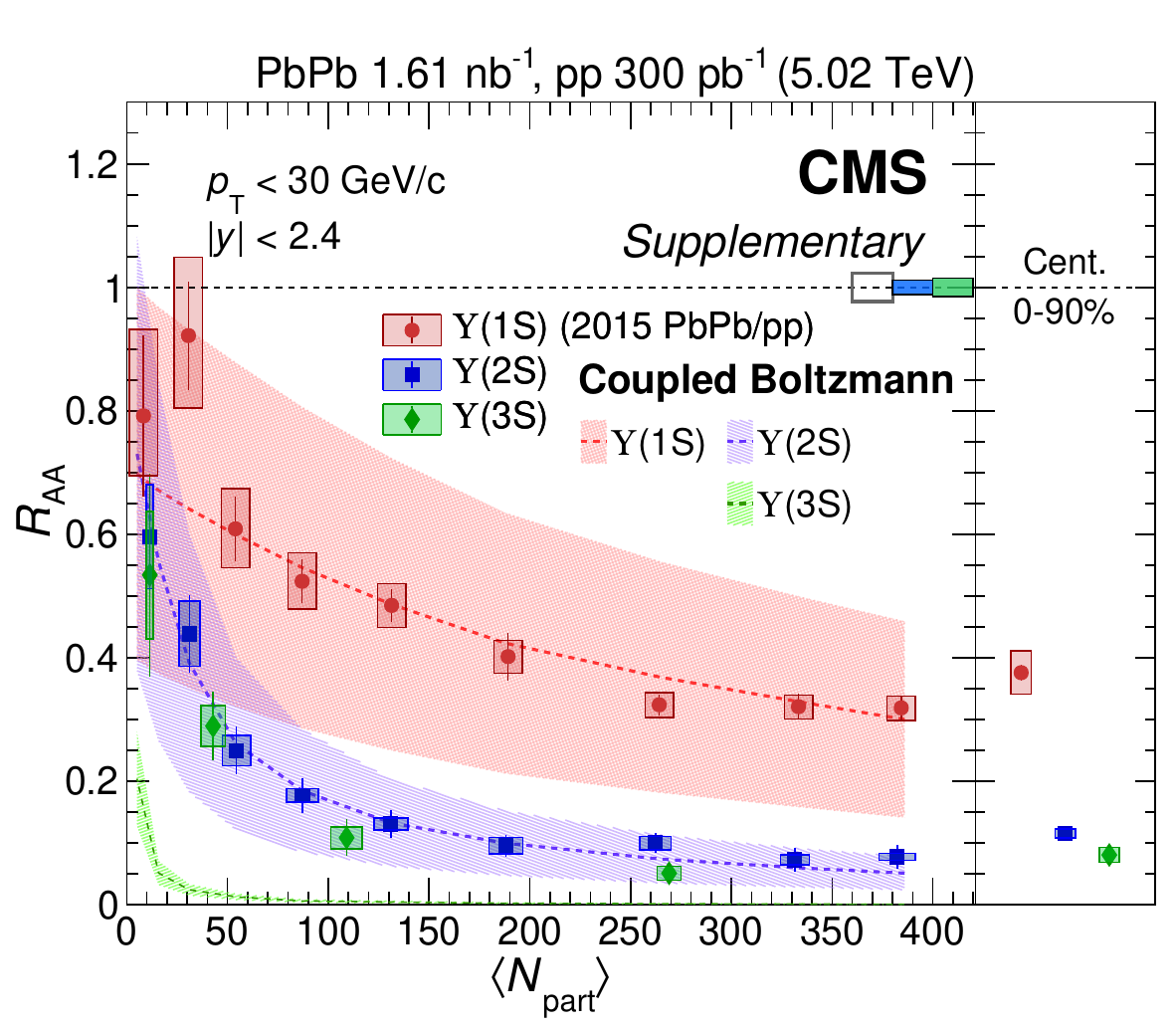}
  \caption{Nuclear modification factors for the \PgUa, \PgUb, and \PgUc mesons as a function of $\langle \Npart \rangle$(from Figure 2 left), including the centrality integrated bin. The vertical lines and boxes correspond to statistical and systematic uncertainties, respectively. The left-most box at unity combines the uncertainties of pp luminosity and \PbPb \NMB, while the second (third) box corresponds to the uncertainty of pp yields for the \PgUb (\PgUc) state. Results for the \PgUa meson are taken from Ref.~\cite{HIN16023}. The bands represent calculations from Ref.~\cite{Xiaojun:2020}.
  }
  \label{fig:theory_cent1}
\end{figure}

\begin{figure}[ht!]
\centering
\includegraphics[width=0.48\textwidth]{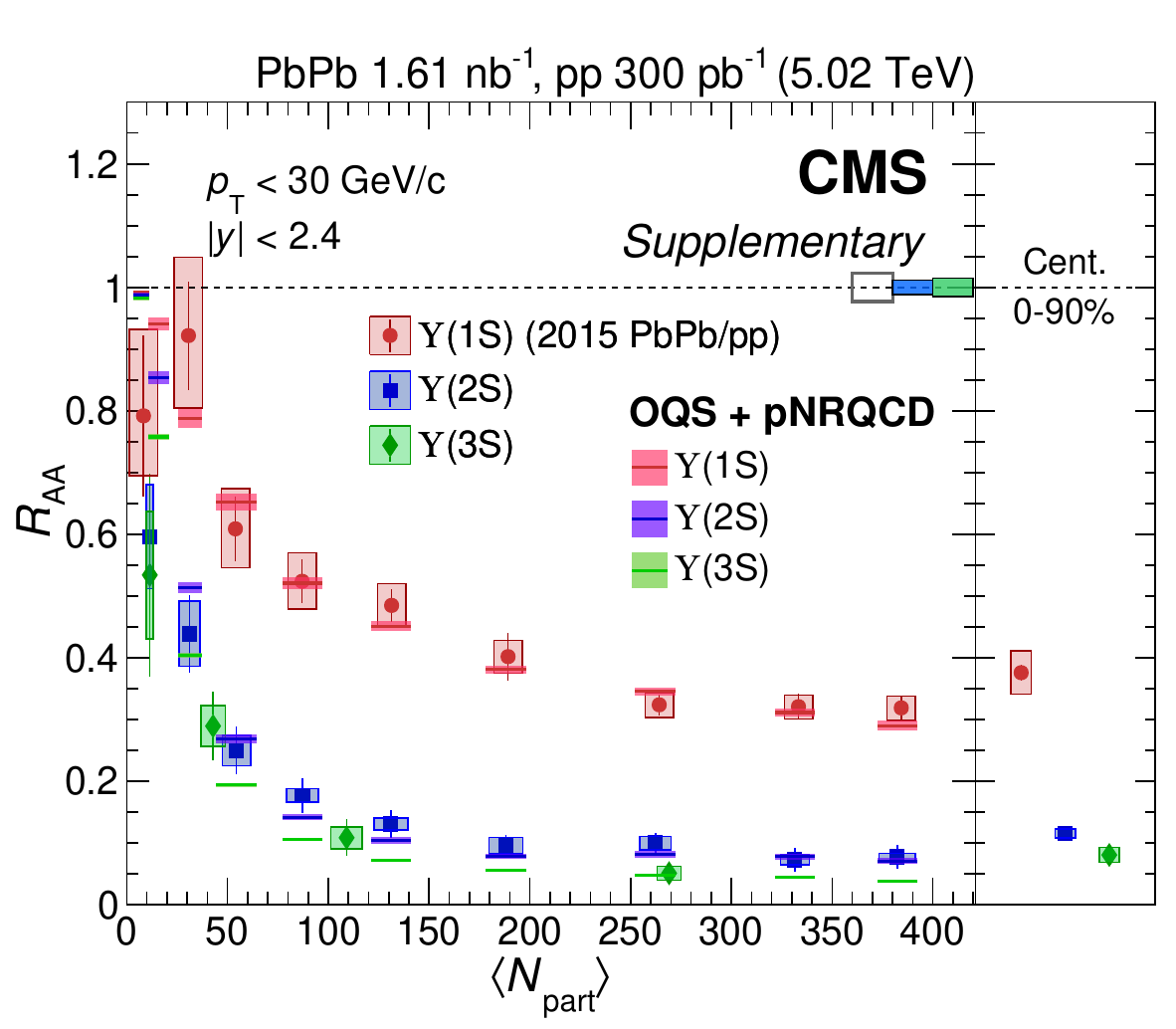}
\caption{Nuclear modification factors for the \PgUa, \PgUb, and \PgUc mesons as a function of $\langle \Npart \rangle$(from Figure 2 left), including the centrality integrated bin. The vertical lines and boxes correspond to statistical and systematic uncertainties, respectively. The left-most box at unity combines the uncertainties of pp luminosity and \PbPb \NMB, while the second (third) box corresponds to the uncertainty of pp yields for the \PgUb (\PgUc) state. Results for the \PgUa meson are taken from Ref.~\cite{HIN16023}. The OQS + pNRQCD theory calculations are taken from Ref.~\cite{Brambilla:2023Reg}.
    }
  \label{fig:theory_cent2}
\end{figure}

\begin{figure}[ht!]
\centering
\includegraphics[width=0.48\textwidth]{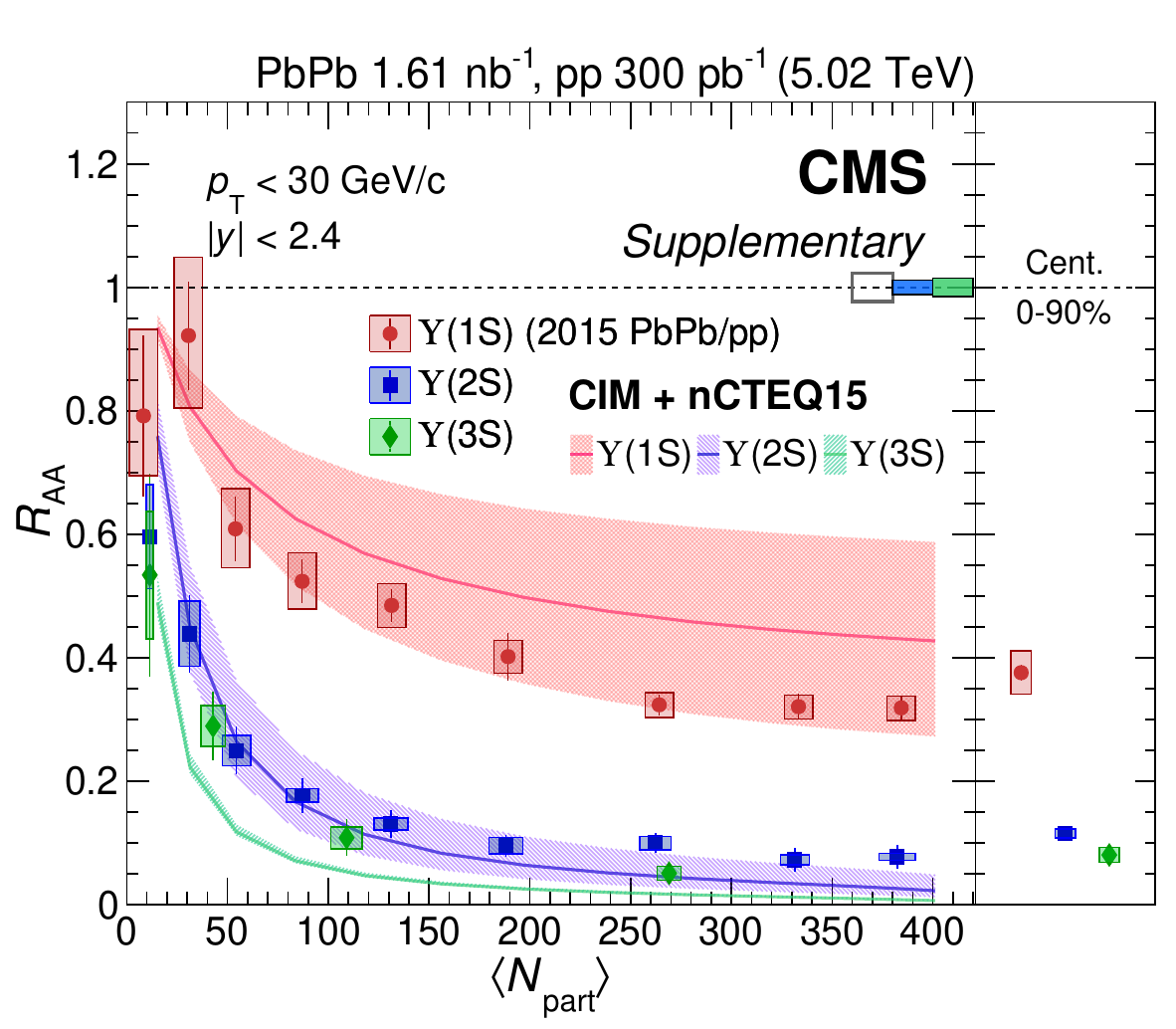}
\caption{Nuclear modification factors for the \PgUa, \PgUb, and \PgUc mesons as a function of $\langle \Npart \rangle$(from Figure 2 left), including the centrality integrated bin. The vertical lines and boxes correspond to statistical and systematic uncertainties, respectively. The left-most box at unity combines the uncertainties of pp luminosity and \PbPb \NMB, while the second (third) box corresponds to the uncertainty of pp yields for the \PgUb (\PgUc) state. Results for the \PgUa meson are taken from Ref.~\cite{HIN16023}. The bands represent calculations from Ref.~\cite{FerreiroComp}.
    }
  \label{fig:theory_cent3}
\end{figure}

\begin{figure}[ht!]
\centering
\includegraphics[width=0.48\textwidth]{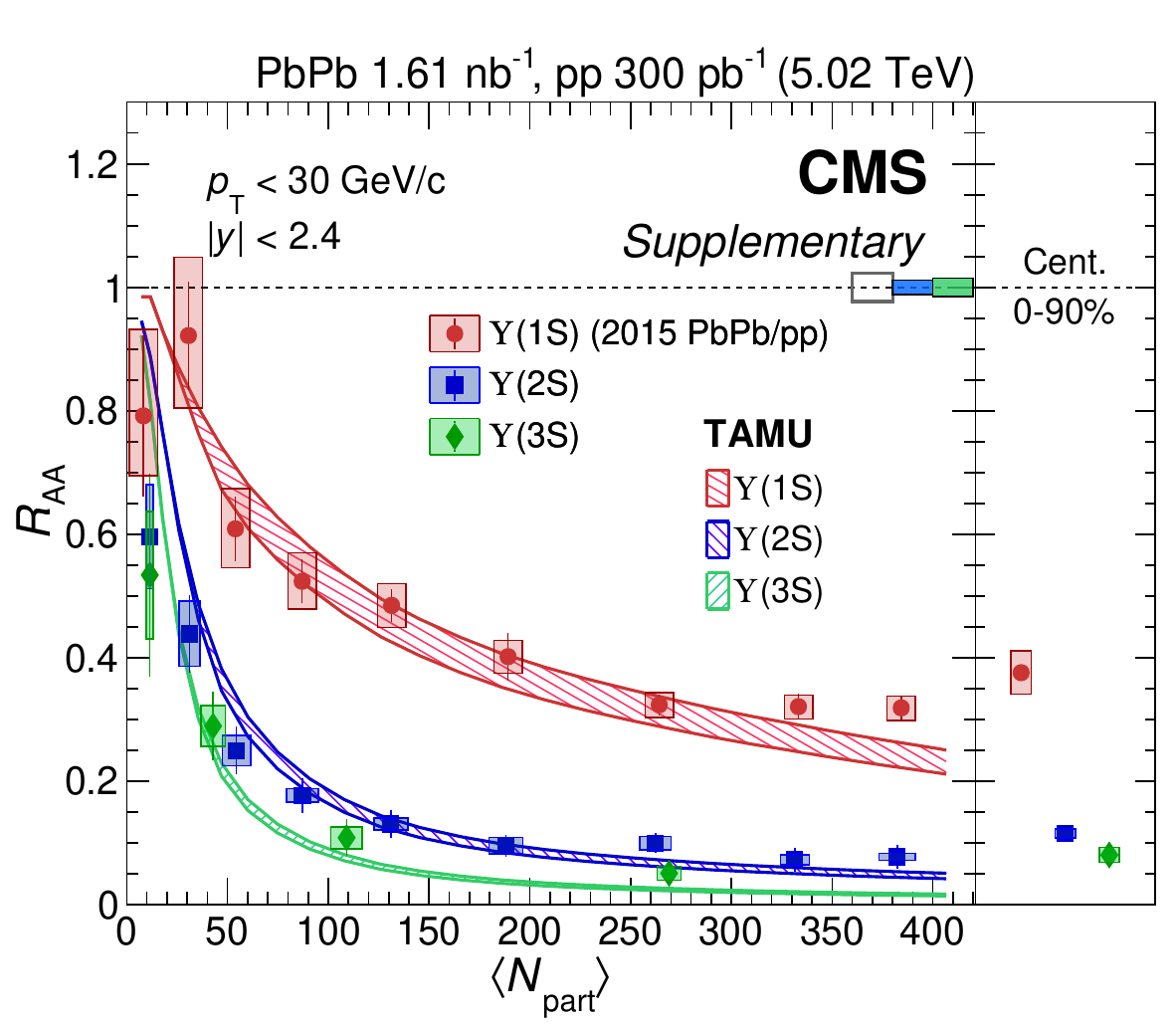}
\caption{Nuclear modification factors for the \PgUa, \PgUb, and \PgUc mesons as a function of $\langle \Npart \rangle$(from Figure 2 left), including the centrality integrated bin. The vertical lines and boxes correspond to statistical and systematic uncertainties, respectively. The left-most box at unity combines the uncertainties of pp luminosity and \PbPb \NMB, while the second (third) box corresponds to the uncertainty of pp yields for the \PgUb (\PgUc) state. Results for the \PgUa meson are taken from Ref.~\cite{HIN16023}. The bands represent calculations from Ref.~\cite{Du:2017qkv}.
    }
  \label{fig:theory_cent4}
\end{figure}

\begin{figure}[ht!]
\centering
\includegraphics[width=0.48\textwidth]{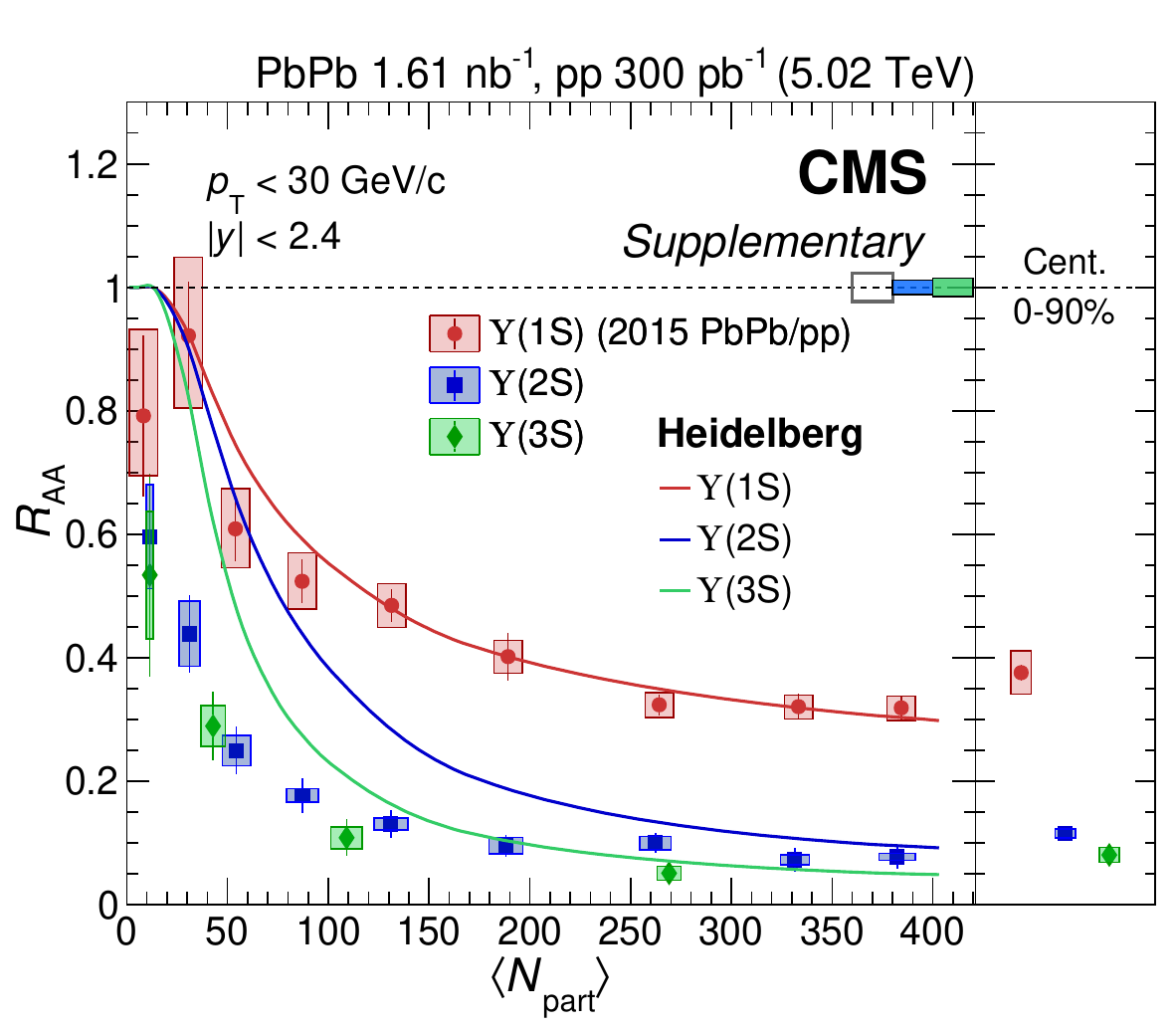}
\caption{Nuclear modification factors for the \PgUa, \PgUb, and \PgUc mesons as a function of $\langle \Npart \rangle$(from Figure 2 left), including the centrality integrated bin. The vertical lines and boxes correspond to statistical and systematic uncertainties, respectively. The left-most box at unity combines the uncertainties of pp luminosity and \PbPb \NMB, while the second (third) box corresponds to the uncertainty of pp yields for the \PgUb (\PgUc) state. Results for the \PgUa meson are taken from Ref.~\cite{HIN16023}. The lines represent calculations from Ref.~\cite{Wolschin:2020ijmpa}.}
\label{fig:theory_cent5}
\end{figure}

\begin{figure}[ht!]
\centering
\includegraphics[width=0.48\textwidth]{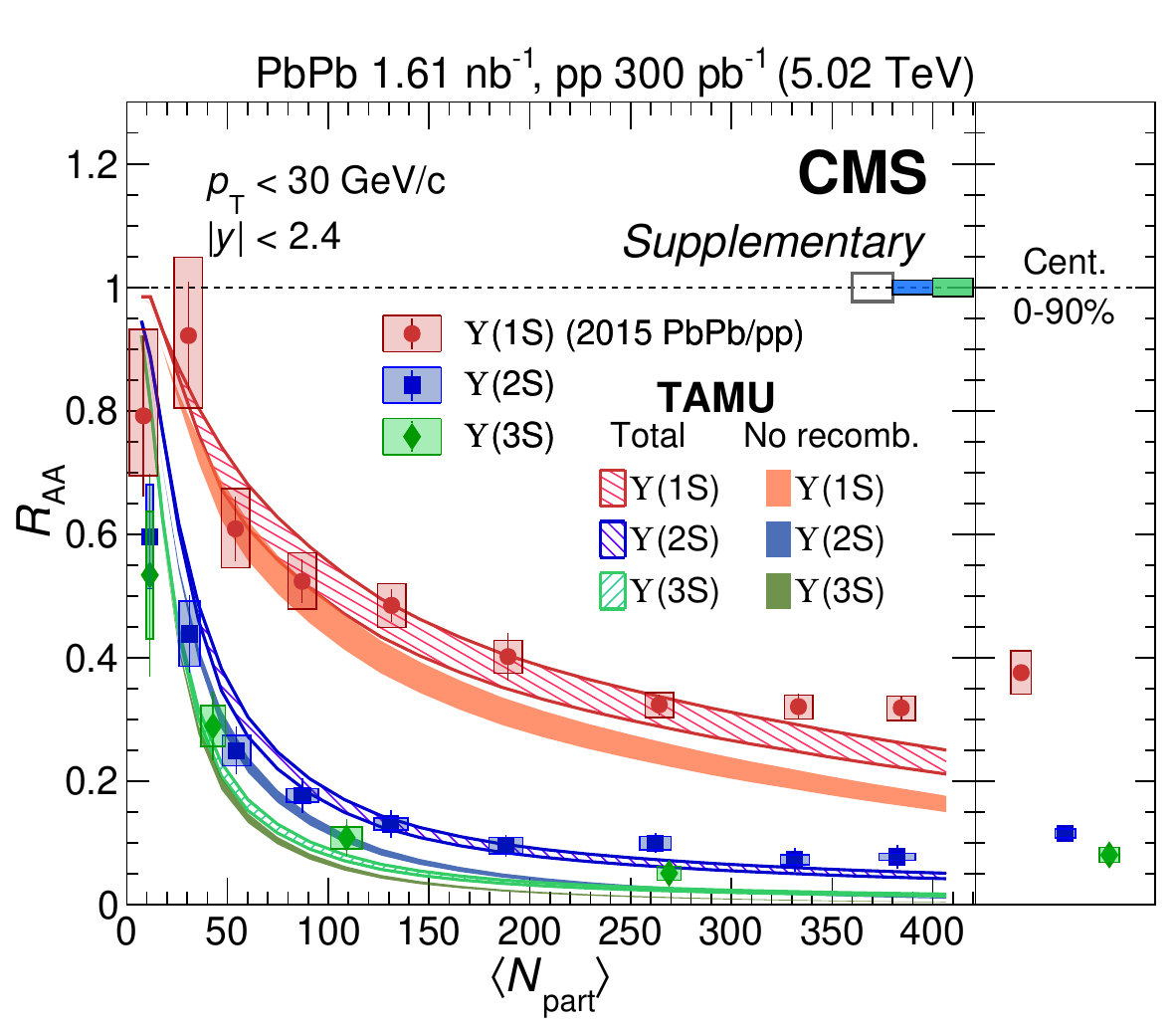}
\caption{Nuclear modification factors for the \PgUa, \PgUb, and \PgUc mesons as a function of $\langle \Npart \rangle$(from Figure 2 left), including the centrality integrated bin. The vertical lines and boxes correspond to statistical and systematic uncertainties, respectively. The left-most box at unity combines the uncertainties of pp luminosity and \PbPb \NMB, while the second (third) box corresponds to the uncertainty of pp yields for the \PgUb (\PgUc) state. Results for the \PgUa meson are taken from Ref.~\cite{HIN16023}. The two type of bands represent calculations from Ref.~\cite{Du:2017qkv}, with the solid filled bands calculated without the recombination component.}
\label{fig:theory_cent6}
\end{figure}

\begin{figure}[ht!]
\centering
\includegraphics[width=0.48\textwidth]{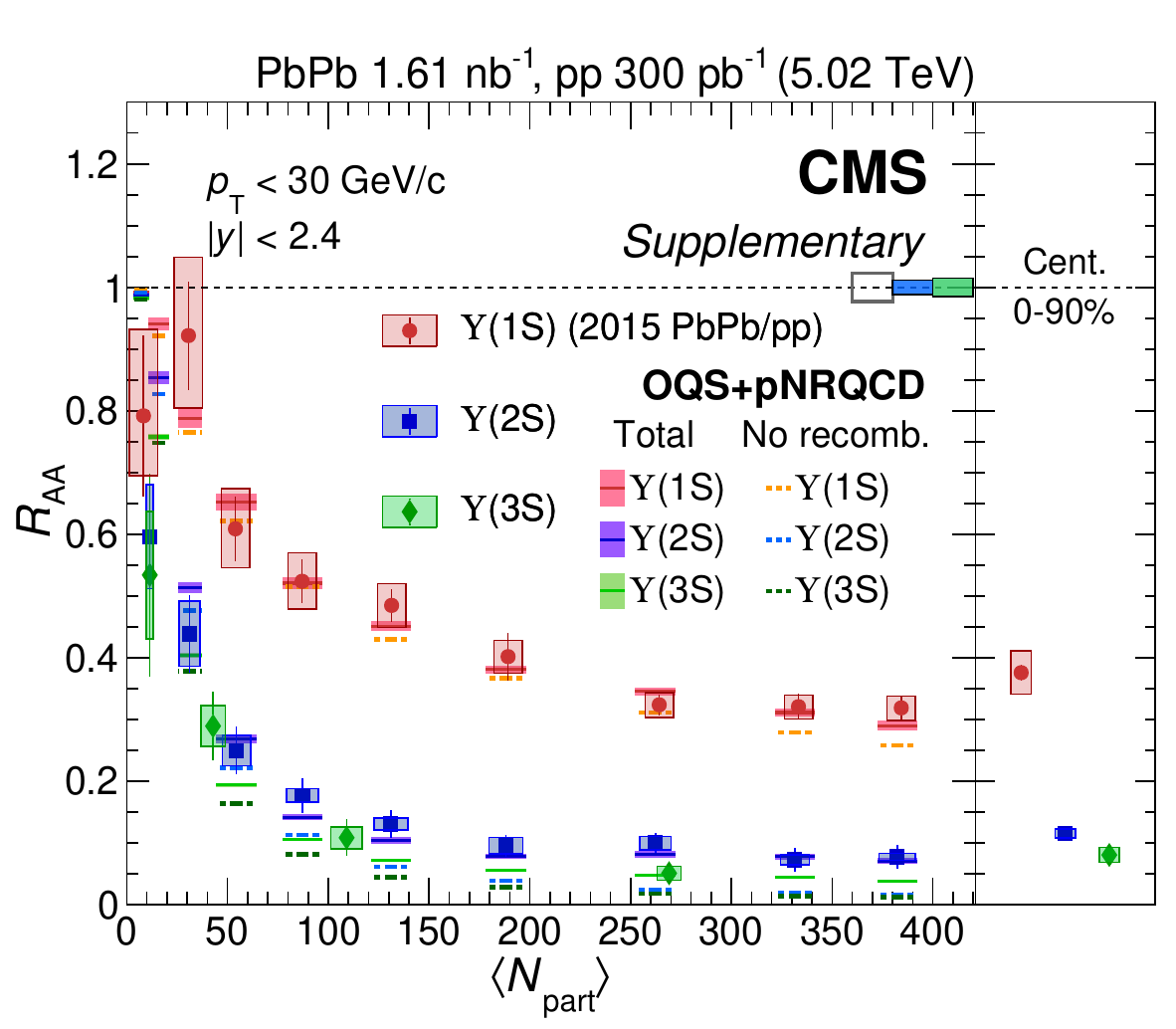}
\caption{Nuclear modification factors for the \PgUa, \PgUb, and \PgUc mesons as a function of $\langle \Npart \rangle$(from Figure 2 left), including the centrality integrated bin. The vertical lines and boxes correspond to statistical and systematic uncertainties, respectively. The left-most box at unity combines the uncertainties of pp luminosity and \PbPb \NMB, while the second (third) box corresponds to the uncertainty of pp yields for the \PgUb (\PgUc) state. Results for the \PgUa meson are taken from Ref.~\cite{HIN16023}. The OQS + pNRQCD theory calculations are taken from Ref.~\cite{Brambilla:2023Reg}, with the dashed one calculated without the recombination component.}
\label{fig:theory_cent7}
\end{figure}

\begin{figure}[ht!]
\centering
\includegraphics[width=0.48\textwidth]{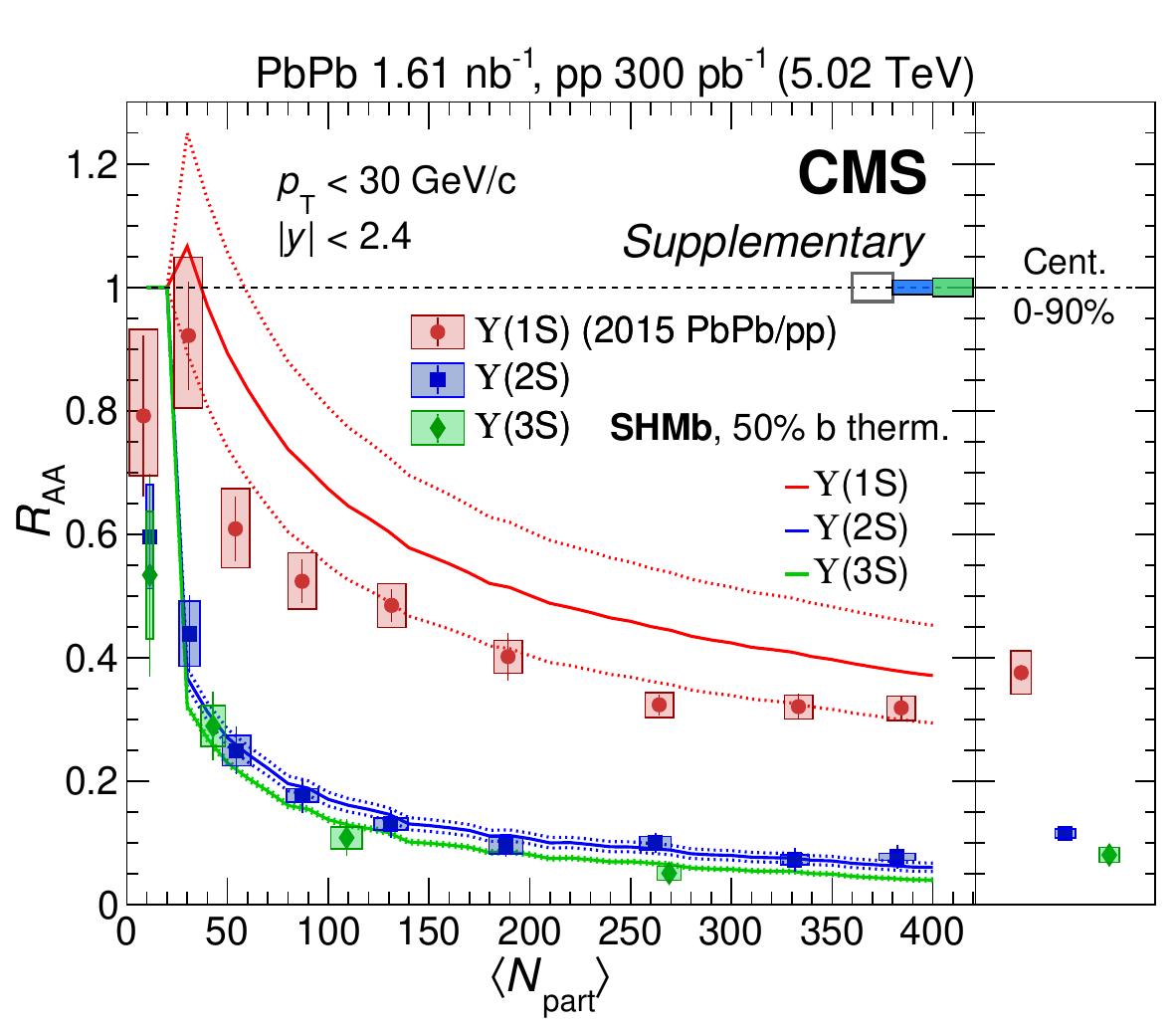}
\caption{Nuclear modification factors for the \PgUa, \PgUb, and \PgUc mesons as a function of $\langle \Npart \rangle$(from Figure 2 left), including the centrality integrated bin. The vertical lines and boxes correspond to statistical and systematic uncertainties, respectively. The left-most box at unity combines the uncertainties of pp luminosity and \PbPb \NMB, while the second (third) box corresponds to the uncertainty of pp yields for the \PgUb (\PgUc) state. Results for the \PgUa meson are taken from Ref.~\cite{HIN16023}. The lines represent calculations from Ref.~\cite{Andronic2018}, with the solid line corresponding to 50\% thermalization of the $b\bar{b}$ pairs and the upper and lower dashed lines representing the 20\% uncertainty of the total $b\bar{b}$ cross section.}
\label{fig:theory_cent8}
\end{figure}

\begin{figure}[ht!]
\centering
\includegraphics[width=0.48\textwidth]{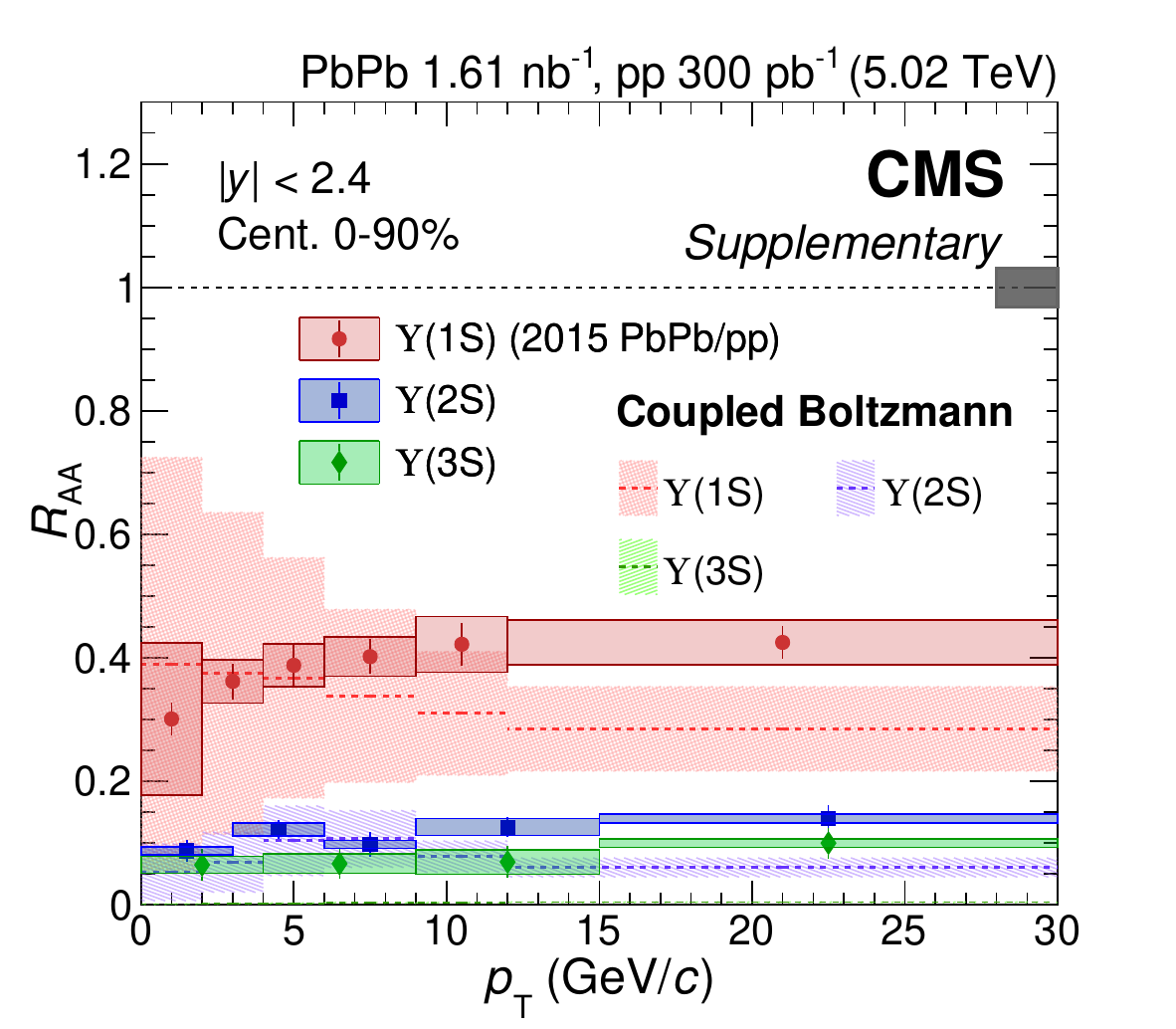}
\caption{Nuclear modification factors for the \PgUa, \PgUb, and \PgUc mesons as a function of \pt(from Figure 2 right). The vertical lines and boxes correspond to statistical and systematic uncertainties, respectively. The box at unity represents the global uncertainty, which combines uncertainties from \TAA, pp luminosity, and \PbPb \NMB. Results for the \PgUa meson are taken from Ref.~\cite{HIN16023}. The bands represent calculations from Ref.~\cite{Xiaojun:2020}.} 
\label{fig:theory_pt1}
\end{figure}

\begin{figure}[ht!]
\centering
\includegraphics[width=0.48\textwidth]{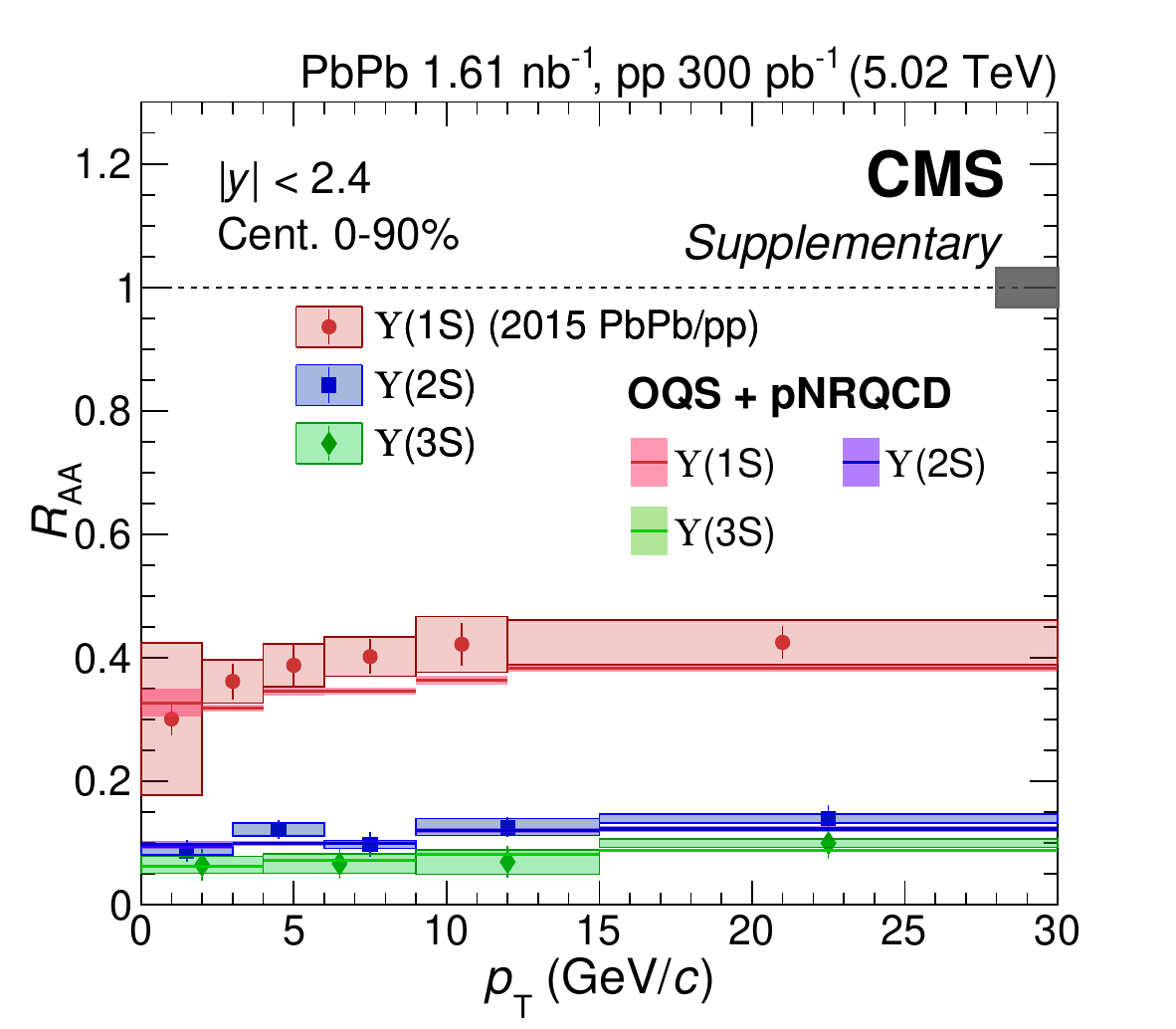}
\caption{Nuclear modification factors for the \PgUa, \PgUb, and \PgUc mesons as a function of \pt(from Figure 2 right). The vertical lines and boxes correspond to statistical and systematic uncertainties, respectively. The box at unity represents the global uncertainty, which combines uncertainties from \TAA, pp luminosity, and \PbPb \NMB. Results for the \PgUa meson are taken from Ref.~\cite{HIN16023}. The OQS + pNRQCD theory calculations are taken from Ref.~\cite{Brambilla:2023Reg}.} 
\label{fig:theory_pt2}
\end{figure}

\begin{figure}[ht!]
\centering
\includegraphics[width=0.48\textwidth]{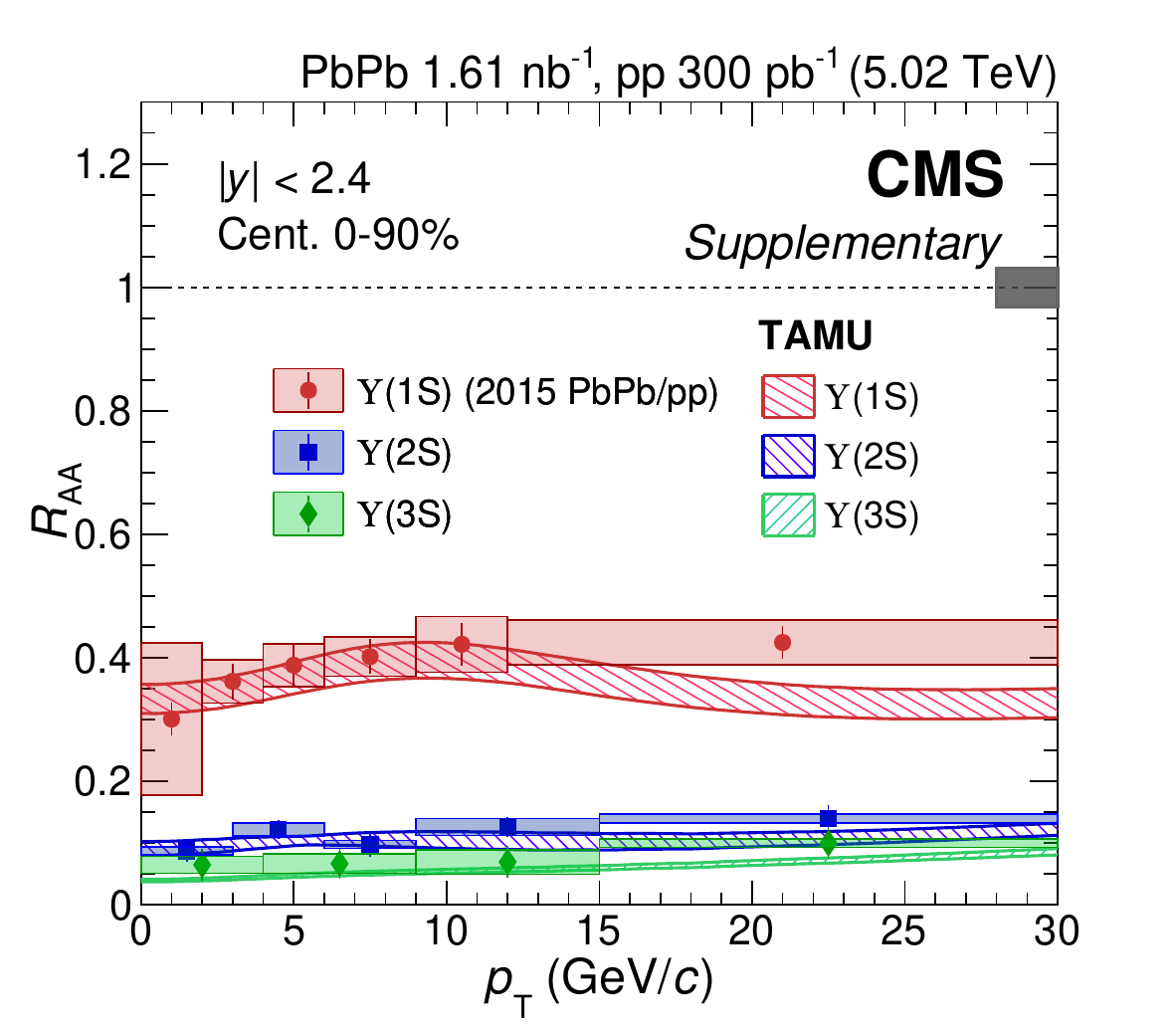}
\caption{Nuclear modification factors for the \PgUa, \PgUb, and \PgUc mesons as a function of \pt(from Figure 2 right). The vertical lines and boxes correspond to statistical and systematic uncertainties, respectively. The box at unity represents the global uncertainty, which combines uncertainties from \TAA, pp luminosity, and \PbPb \NMB. Results for the \PgUa meson are taken from Ref.~\cite{HIN16023}. The bands represent calculations from Ref.~\cite{Du:2017qkv}.} 
\label{fig:theory_pt3}
\end{figure}

\begin{figure}[ht!]
\centering
\includegraphics[width=0.48\textwidth]{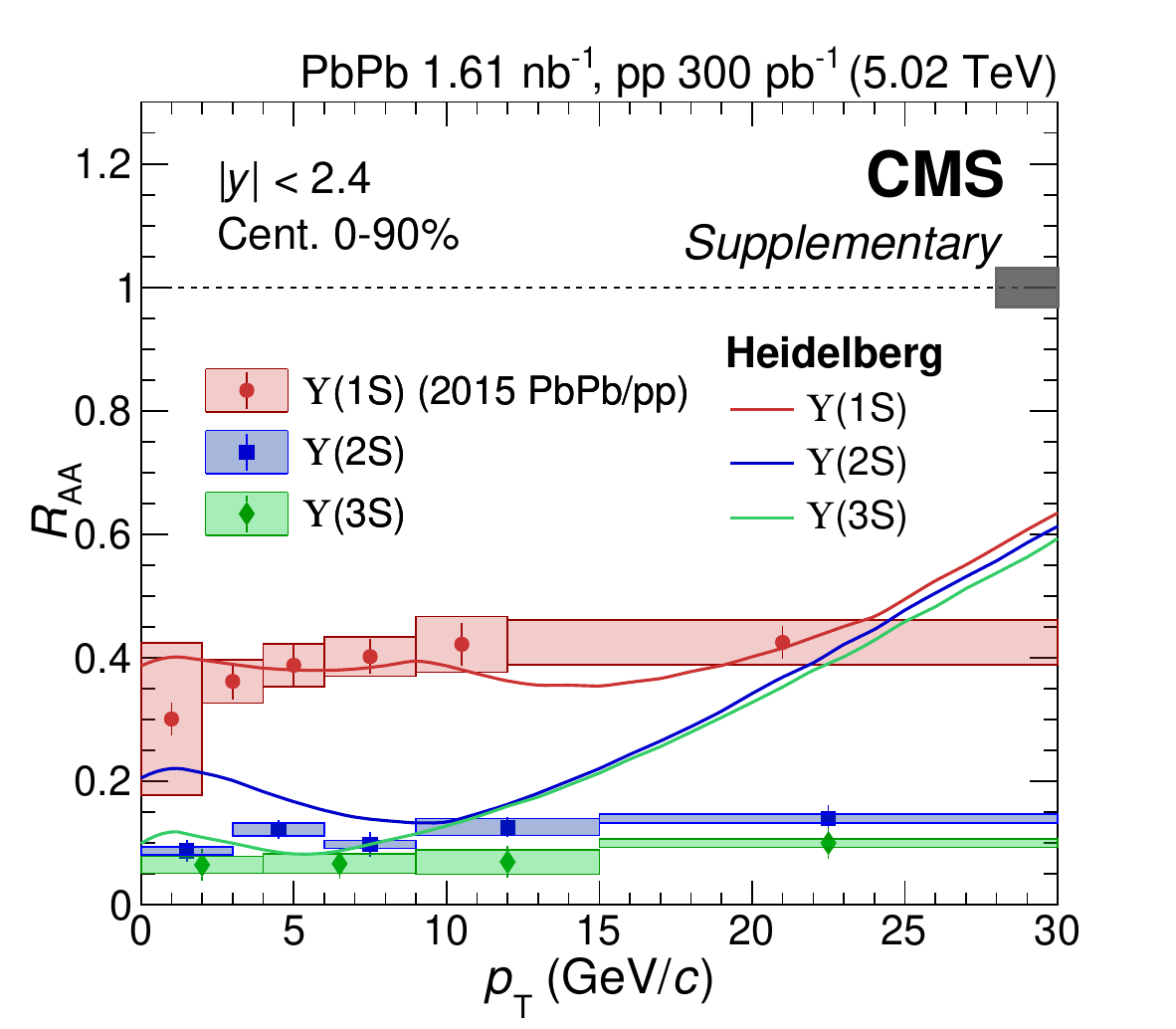}
\caption{Nuclear modification factors for the \PgUa, \PgUb, and \PgUc mesons as a function of \pt(from 2 right). The vertical lines and boxes correspond to statistical and systematic uncertainties, respectively. The box at unity represents the global uncertainty, which combines uncertainties from \TAA, pp luminosity, and \PbPb \NMB. Results for the \PgUa meson are taken from Ref.~\cite{HIN16023}. The lines represent calculations from Ref.~\cite{Wolschin:2020ijmpa}.} 
\label{fig:theory_pt4}
\end{figure}

\begin{figure}[ht!]
\centering
\includegraphics[width=0.48\textwidth]{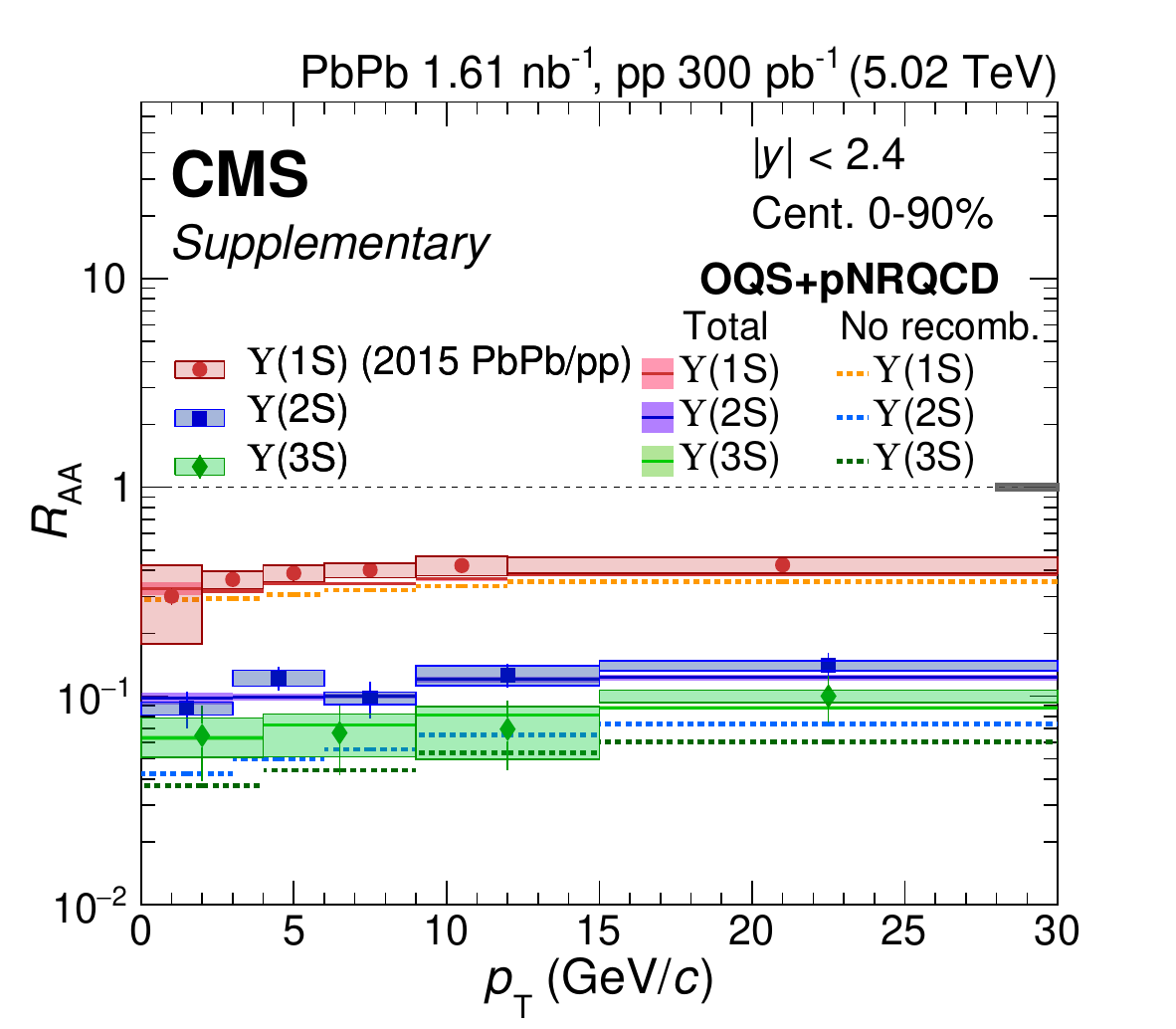}
\caption{Nuclear modification factors for the \PgUa, \PgUb, and \PgUc mesons as a function of \pt(from Figure 2 right). The vertical lines and boxes correspond to statistical and systematic uncertainties, respectively. The box at unity represents the global uncertainty, which combines uncertainties from \TAA, pp luminosity, and \PbPb \NMB. Results for the \PgUa meson are taken from Ref.~\cite{HIN16023}. The OQS + pNRQCD theory calculations are taken from Ref.~\cite{Brambilla:2023Reg}, with the dashed one calculated without the recombination component.}
\label{fig:theory_pt5}
\end{figure}

\begin{figure}[ht!]
\centering
\includegraphics[width=0.48\textwidth]{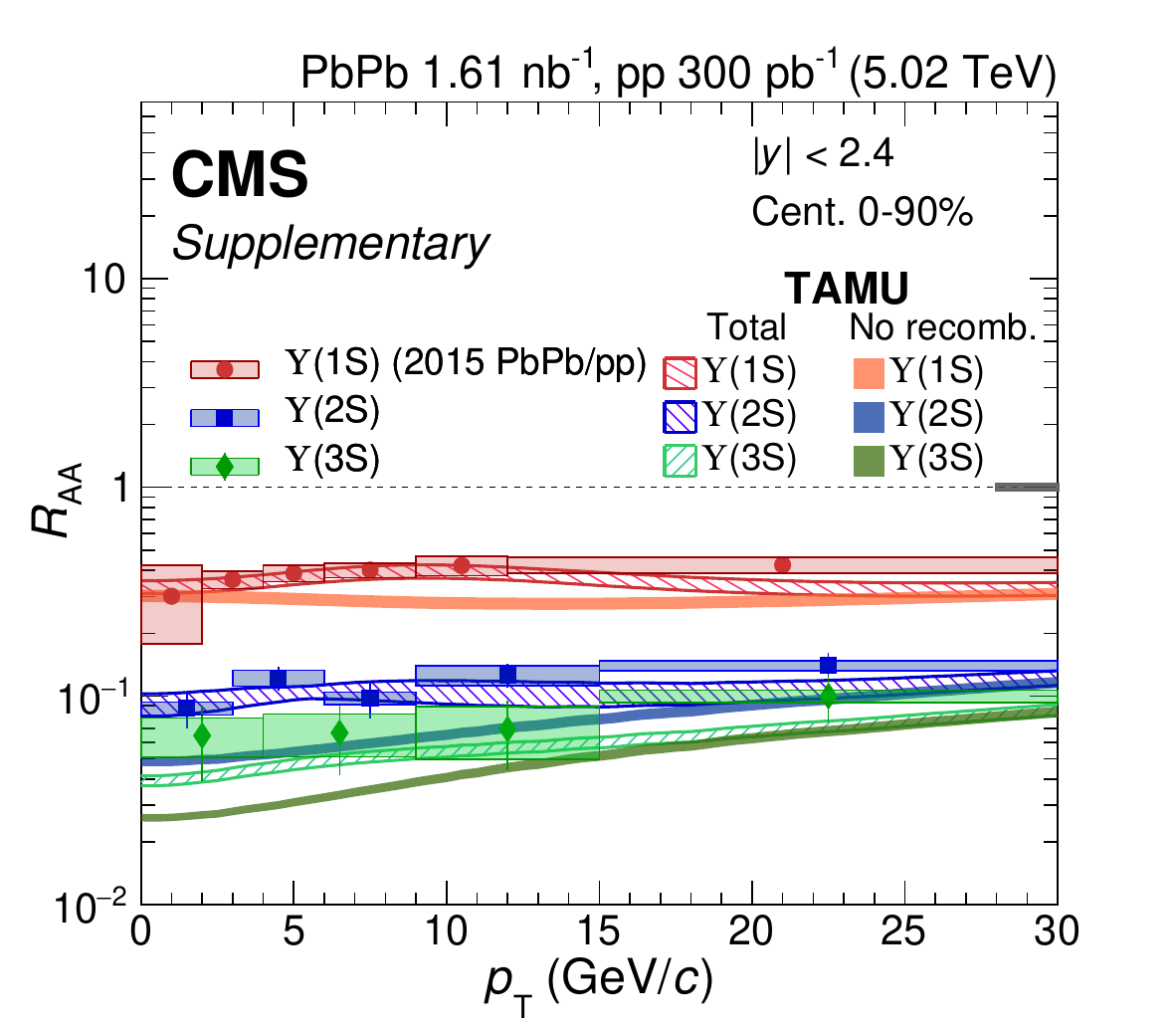}
\caption{Nuclear modification factors for the \PgUa, \PgUb, and \PgUc mesons as a function of \pt(from Figure 2 right). The vertical lines and boxes correspond to statistical and systematic uncertainties, respectively. The box at unity represents the global uncertainty, which combines uncertainties from \TAA, pp luminosity, and \PbPb \NMB. Results for the \PgUa meson are taken from Ref.~\cite{HIN16023}. The two type of bands represent calculations from Ref.~\cite{Du:2017qkv}, with the solid filled bands calculated without the recombination component.} 
\label{fig:theory_pt6}
\end{figure}

\begin{figure}[ht!]
\centering
\includegraphics[width=0.48\textwidth]{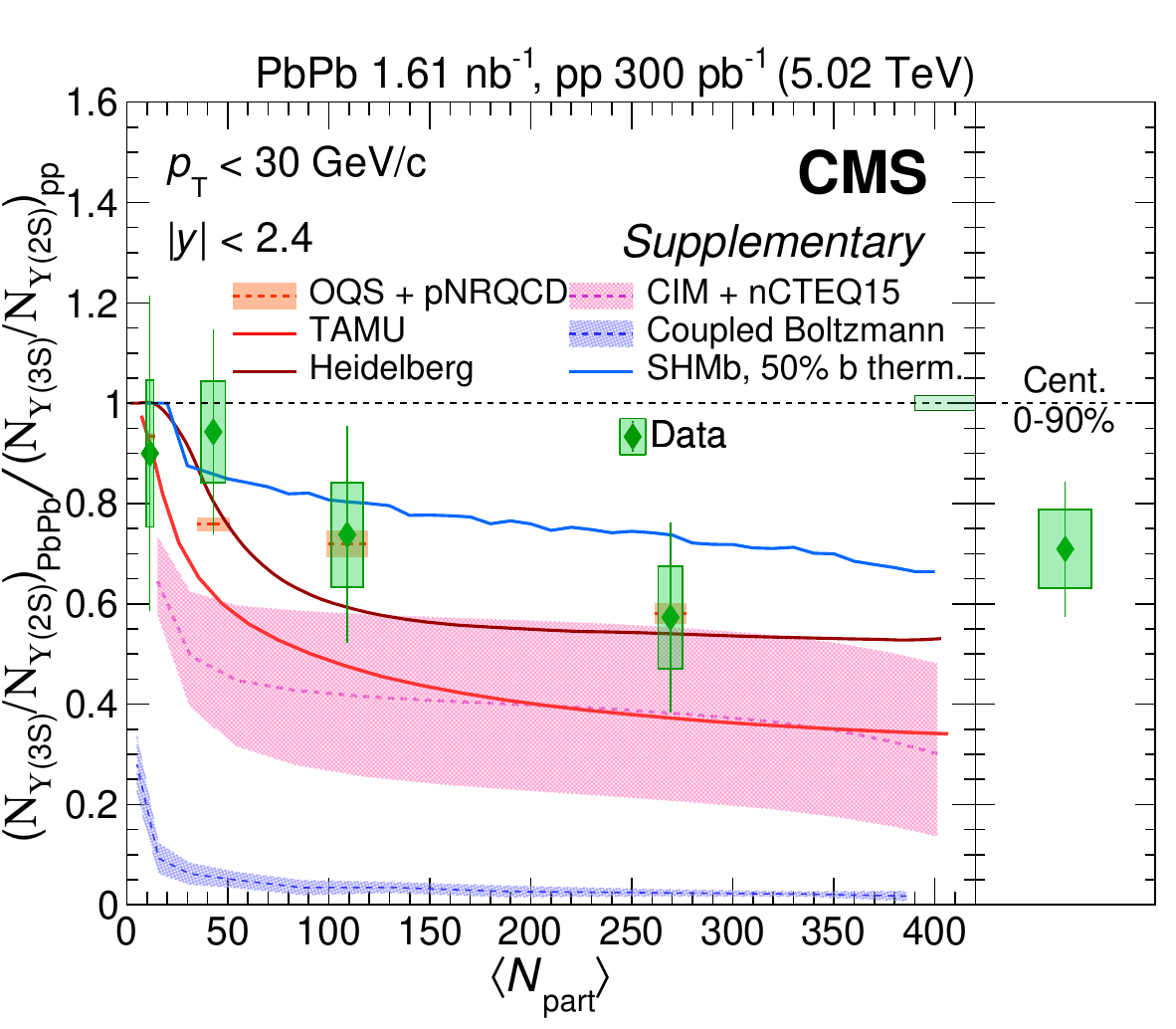}
\caption{The double ratio of \PgUc/\PgUb as functions of $\langle \Npart \rangle$(from Figure 3 left). The vertical lines correspond to statistical uncertainties, while the boxes are the systematic uncertainties. The box at unity shows the combined systematic and statistical uncertainties from pp data. The six different types of lines and bands represent calculations from Ref.~\cite{Xiaojun:2020,Brambilla:2023Reg,FerreiroComp,Du:2017qkv,Wolschin:2020ijmpa,Andronic2018}.}
\label{fig:theory_dr_cent1}
\end{figure}

\begin{figure}[ht!]
\centering
\includegraphics[width=0.48\textwidth]{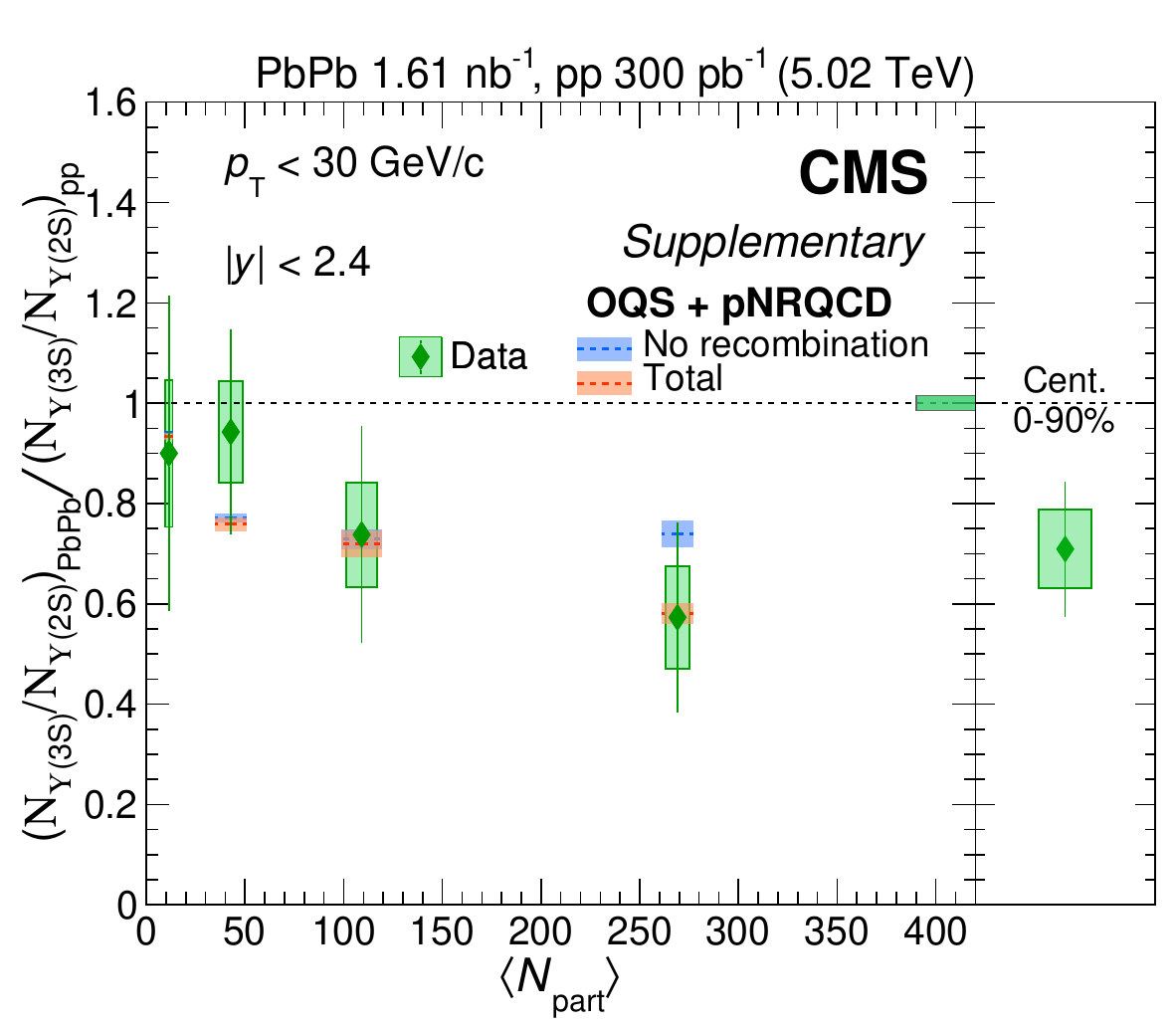}
          \caption{The double ratio of \PgUc/\PgUb as a function of $\langle \Npart \rangle$(from Figure 3 left). The vertical lines correspond to statistical uncertainties, while the boxes are the systematic uncertainties. The box at unity shows the combined systematic and statistical uncertainties from pp data. The orange and blue boxes represent calculations from Ref.~\cite{Brambilla:2023Reg}, with the latter showing the calculations without the recombination component.}
\label{fig:theory_dr_cent2}
\end{figure}

\begin{figure}[ht!]
\centering
\includegraphics[width=0.48\textwidth]{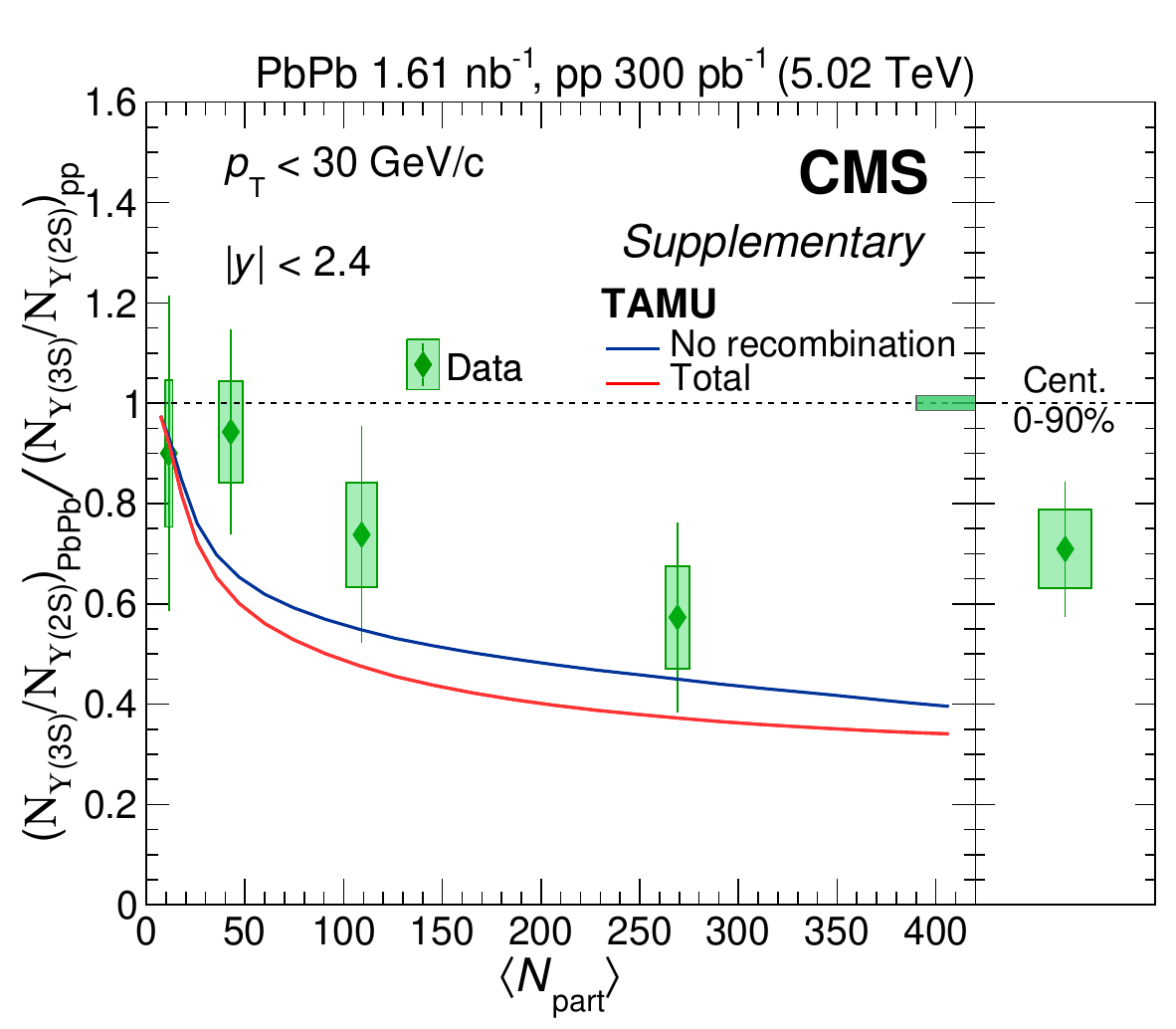}
          \caption{The double ratio of \PgUc/\PgUb as a function of $\langle \Npart \rangle$(from Figure 3 left). The vertical lines correspond to statistical uncertainties, while the boxes are the systematic uncertainties. The box at unity shows the combined systematic and statistical uncertainties from pp data. The red and blue lines represent calculations from Ref.~\cite{Du:2017qkv}, with the latter showing the calculations without the recombination component.}
\label{fig:theory_dr_cent3}
\end{figure}

\begin{figure}[ht!]
\centering
\includegraphics[width=0.48\textwidth]{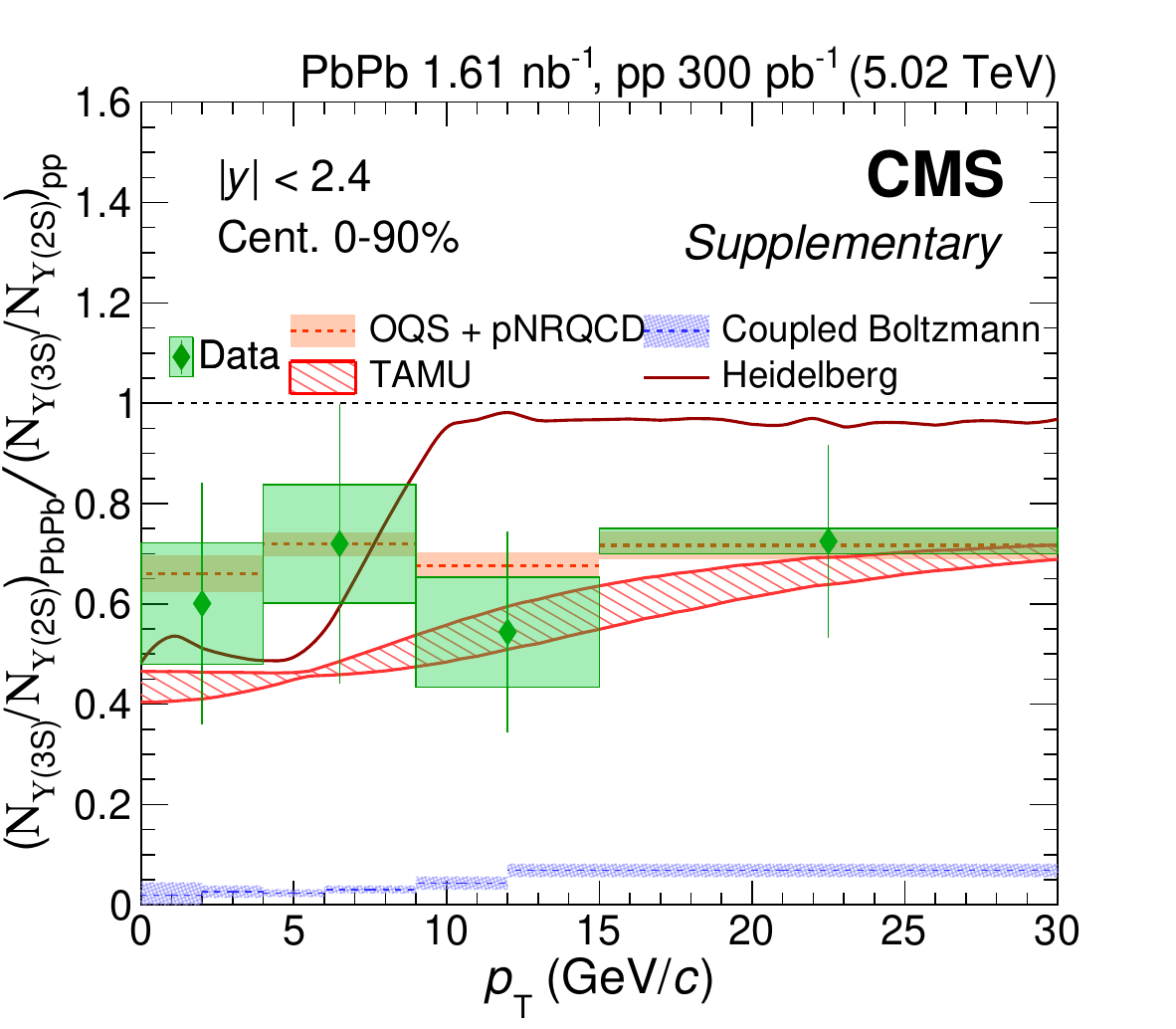}
          \caption{The double ratio of \PgUc/\PgUb as a function of \pt(from Figure 3 right). The vertical lines correspond to statistical uncertainties, while the boxes are the systematic uncertainties. The bands and line represent calculations from Ref.~\cite{Xiaojun:2020,Brambilla:2023Reg,Du:2017qkv,Wolschin:2020ijmpa}.}
\label{fig:theory_dr_pt1}
\end{figure}

\begin{figure}[ht!]
\centering
\includegraphics[width=0.48\textwidth]{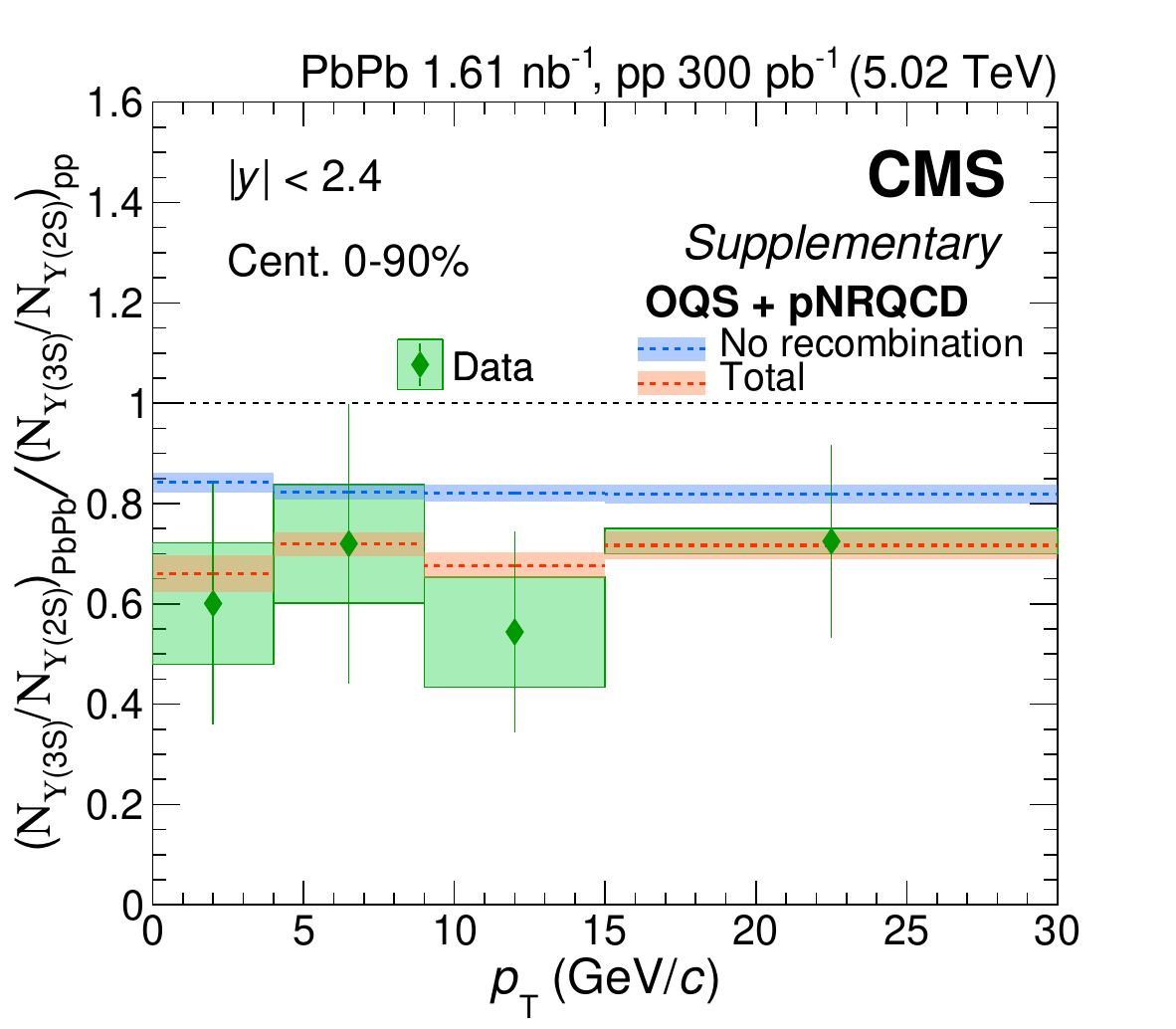}
          \caption{The double ratio of \PgUc/\PgUb as a function of \pt(from Figure 3 right). The vertical lines correspond to statistical uncertainties, while the boxes are the systematic uncertainties. The orange and blue bands represent calculations from Ref.~\cite{Brambilla:2023Reg}, with the latter showing the calculations without the recombination component.}
\label{fig:theory_dr_pt2}
\end{figure}

\begin{figure}[ht!]
\centering
\includegraphics[width=0.48\textwidth]{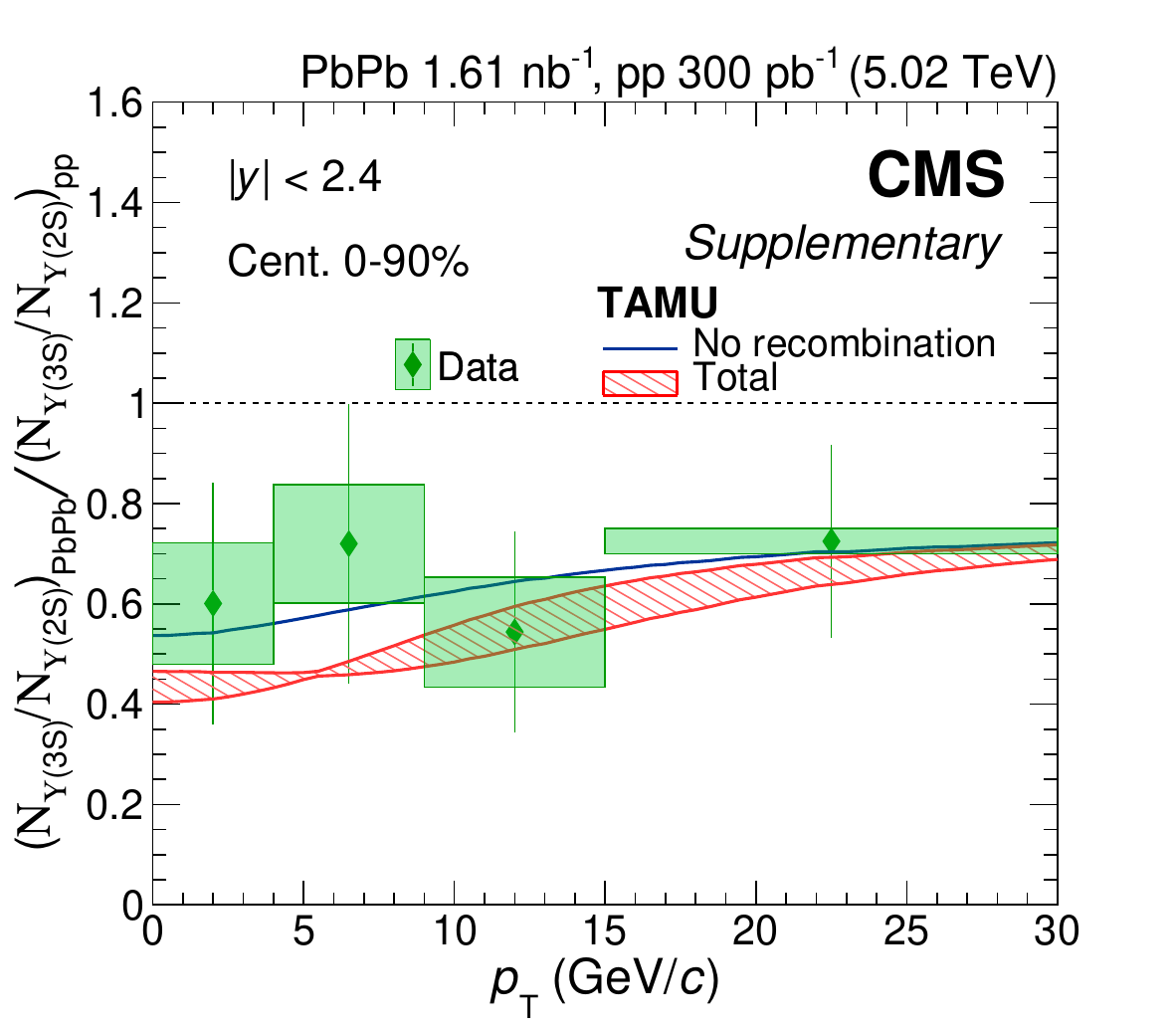}
          \caption{The double ratio of \PgUc/\PgUb as a function of \pt(from Figure 3 right). The vertical lines correspond to statistical uncertainties, while the boxes are the systematic uncertainties. The red band and blue line represent calculations from Ref.~\cite{Du:2017qkv}, with the latter showing the calculations without the recombination component.}
\label{fig:theory_dr_pt3}
\end{figure}

\begin{figure}[ht!]
\centering
\includegraphics[width=0.53\textwidth]{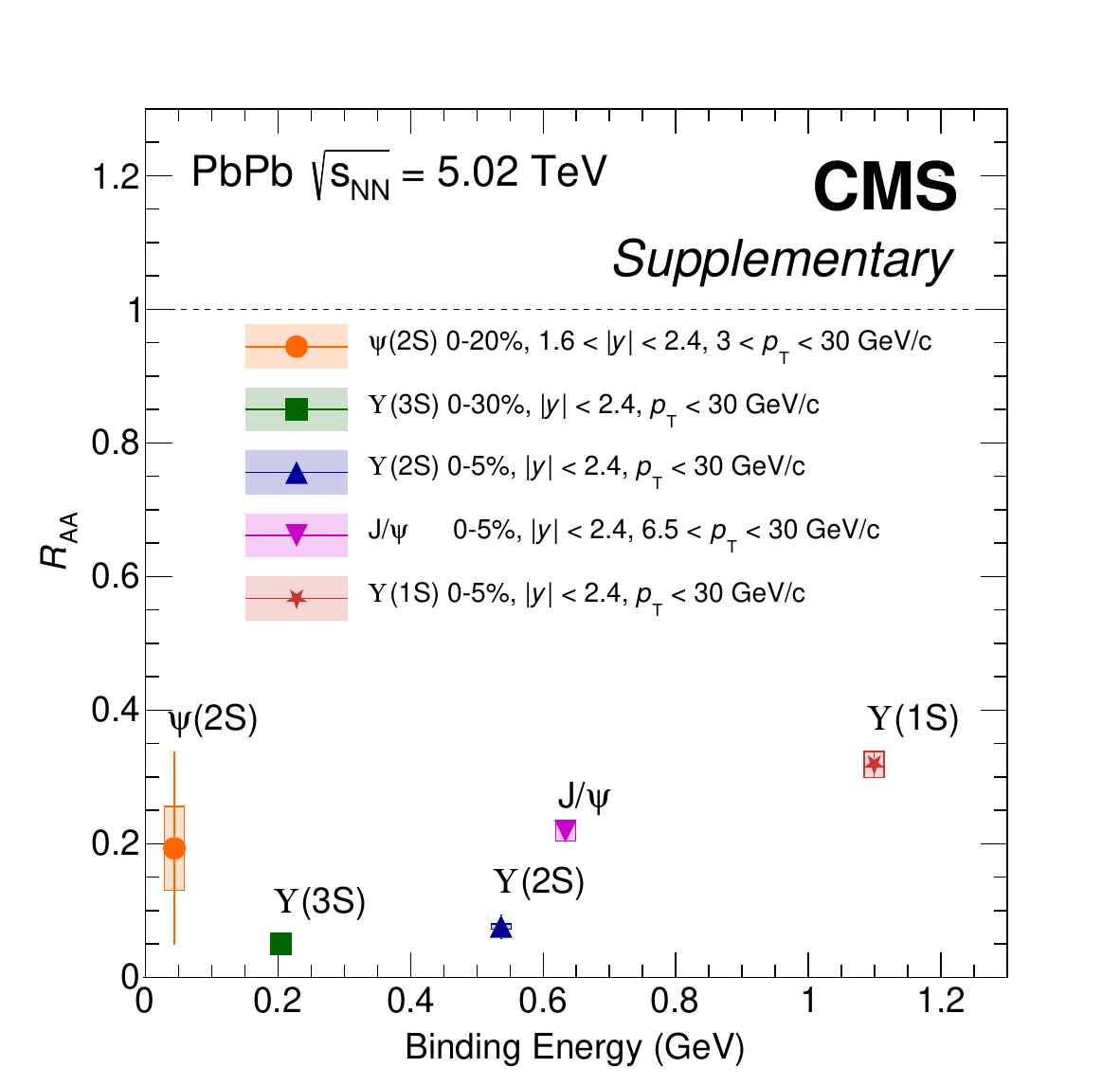}
          \caption{The nuclear modification factors for various quarkonium mesons as a function of quarkonium binding energy at $\sqrt{\mathrm{s_{NN}}}=5.02~\mathrm{TeV}$. The \RAA values are taken from the data point with the most central collision bin, e.g., the values for the \PgU's correspond to the points of the hightest \Npart in Figure 2 left. The values for the binding energy of each quarkonium state are taken from Ref.~\cite{BindingESatz}. The error bars and boxes represent the statistical and systematic uncertainties, respectively. The results for the \PgUa meson and charmonium states (\PJGy and \Pgy) are taken from Refs.~\cite{HIN16023} and~\cite{Sirunyan:2018mp}, respectively.}
\label{fig:raa_bindingE}
\end{figure}

\begin{figure}[ht!]
\centering
\includegraphics[width=0.48\textwidth]{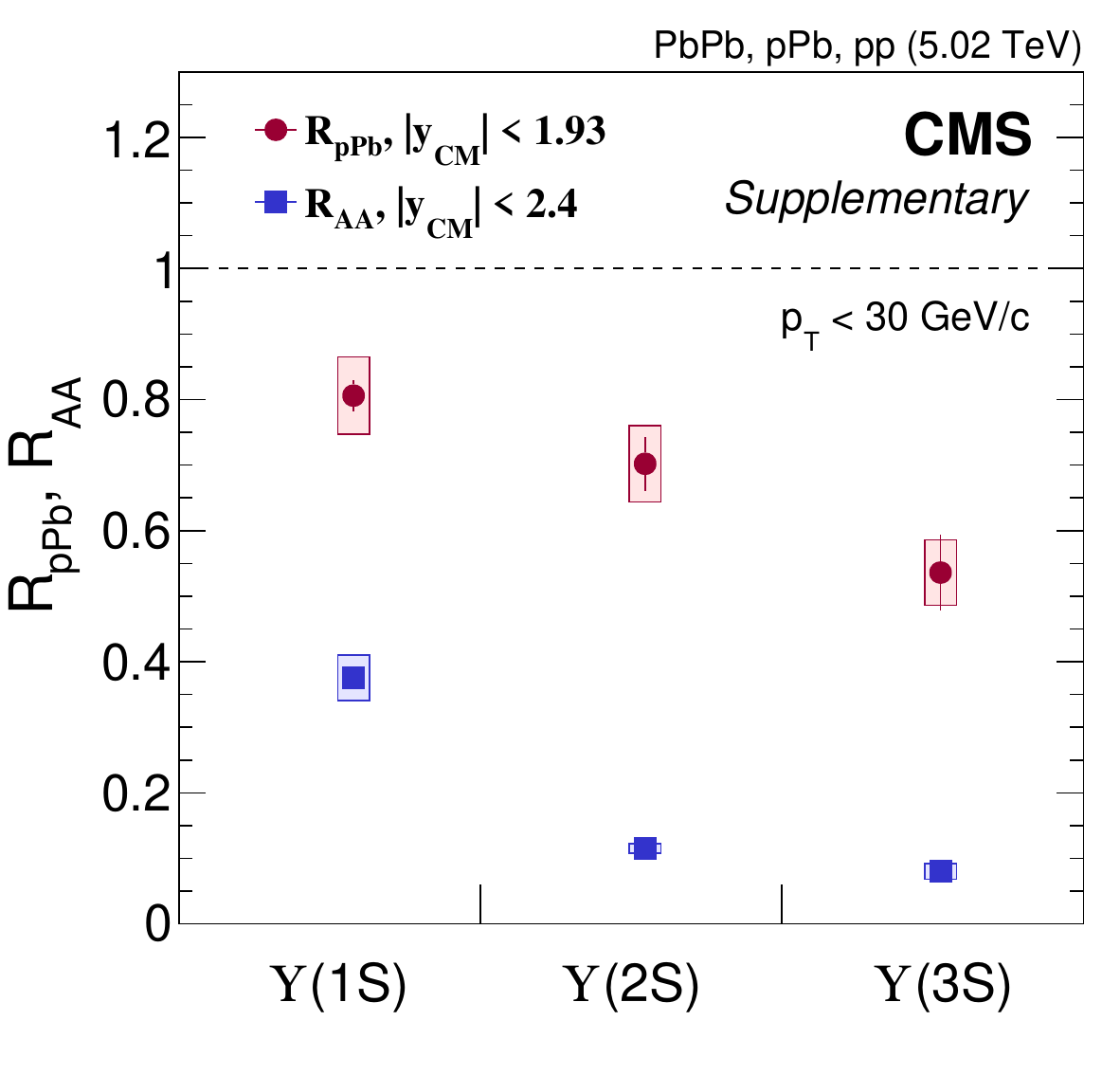}
          \caption{The nuclear modification factors for \PgUa, \PgUb, and \PgUc mesons in pPb and \PbPb collisions at $\sqrt{\mathrm{s_{NN}}}=5.02~\mathrm{TeV}$. The error bars and boxes represent the statistical and systematic uncertainties, respectively. The \PgUb and \PgUc values are from the integrated bin in Figure 2 left. The results for pPb collisions and the \PgUa meson are taken from Refs.~\cite{HIN18005} and~\cite{HIN16023}, respectively.} 
\label{fig:raa_rpa}
\end{figure}

\begin{figure*}[ht!]
\centering
\includegraphics[width=0.7\textwidth]{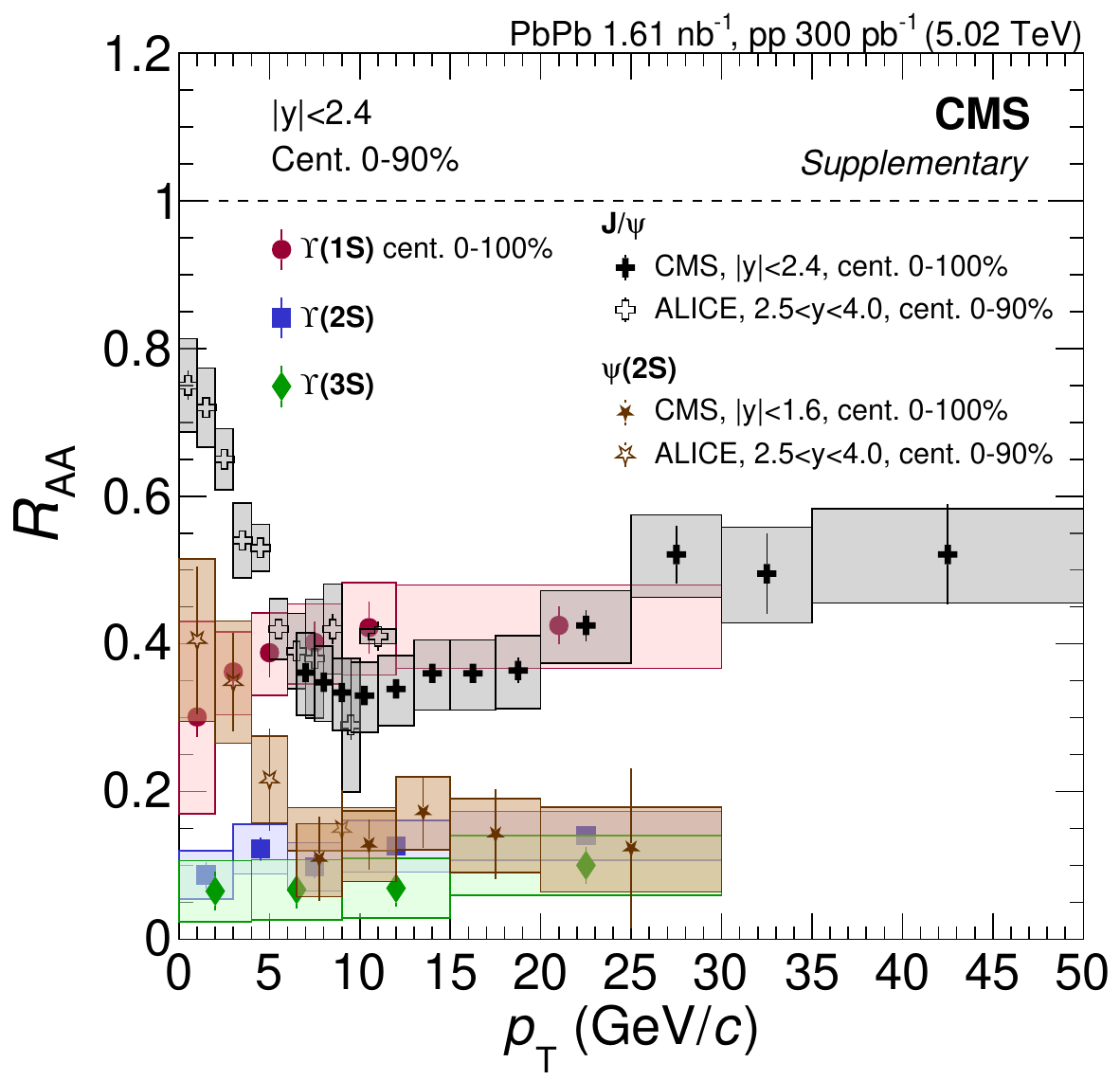}
  \caption{Nuclear modification factors for the \PgUa, \PgUb, \PgUc, \PJGy, and \Pgy mesons as a function of \pt. The vertical lines and boxes correspond to statistical and systematic uncertainties, respectively. Results for the \PgUa meson are taken from Ref.~\cite{HIN16023}. The open and full cross points are the results for \PJGy mesons from Refs.~\cite{ALICE:JpsiRAA2019} and~\cite{Sirunyan:2018mp}, respectively. Results for \Pgy mesons are taken from Refs.~\cite{ALICE:2022psi2sRAA} and~\cite{Sirunyan:2018mp} for the open and full star points, respectively.}
\label{fig:raa_pt_exp}
\end{figure*}

\begin{figure}[h!]
\centering
      \includegraphics[width=0.48\textwidth]{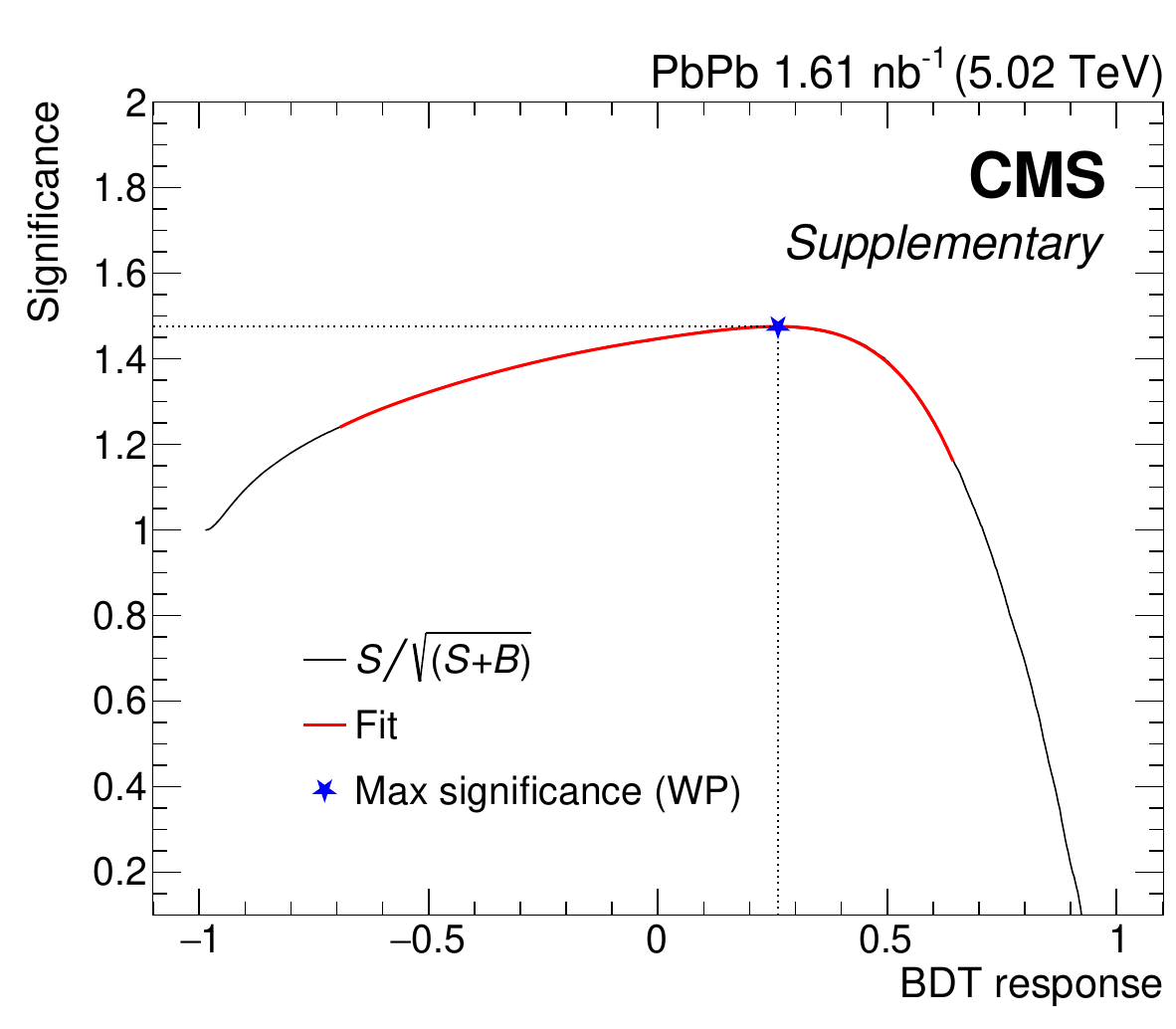}
      \caption{The significance $S/\sqrt{S+B}$ for \PgU mesons in \PbPb collisions as a function of BDT score normalized to range between -1 and 1. The quantities $S$ and $B$ represent the yields of signal and background dimuons used for the BDT training, respectively. The working point (WP) is determined to be the BDT score that maximizes the significance using a parametic fit.}
\label{fig:bdt_signif}
\end{figure}

\clearpage

\begin{table}[!hp]
  \centering
    \begin{tabular}{cccc}
      \pt range (\GeVc) & $\frac{1}{\langle \TAA \rangle}\frac{\rd^2 N^{\PgU}}{\rd\pt\rd y}$ & Stat. Unc.               & Syst. Unc.           \\
      \hline
      0--3                                             & 2.98            & 0.60 & 0.22  \\
      3--6                                             & 5.51            & 0.72 & 0.48  \\
      6--9                                             & 2.72            & 0.54 & 0.19  \\
      9--15                                            & 1.17            & 0.15 & 0.13  \\
      15--30                                           & 0.15            & 0.02 & 0.01 
    \end{tabular}
    \caption{Yields for \PgUb mesons in \PbPb collisions in centrality 0--90\% and $\abs{y} < 2.4$, corrected for acceptance and efficiency, and normalized by the nuclear thickness function $\langle \TAA \rangle$ and the number of minimum bias events $\NMB$. The values for the yields and their uncertainties are in units of pb$/\GeVc$.}
    \label{table:yield_pt_2S}
\end{table}
      
\begin{table}[!hp]
  \centering
    \begin{tabular}{ccccc}
      Centrality & $\langle \Npart \rangle$ & $\frac{1}{\langle \TAA \rangle}\frac{\rd^2 N^{\PgU}}{\rd\pt\rd y}$ & Stat. Unc. & Syst. Unc.           \\
      \hline
      0--5\% &  382.30    &  1.02 &  0.25 & 0.08  \\
      5--10\% &  331.50   &  0.95 &  0.25 & 0.11 \\
      10--20\% &  262.30  &  1.31 &  0.20 & 0.14  \\
      20--30\% &  188.20  &  1.25 &  0.24 & 0.17  \\
      30--40\% &  131.00  &  1.71 &  0.29 & 0.13  \\
      40--50\% &  87.19  &  2.32 &  0.37 &  0.15  \\
      50--60\% &  54.42  &  3.27 &  0.51 &  0.32  \\
      60--70\% &  31.21  &  5.74 &  0.82 &  0.69  \\
      70--90\% &  11.43  &  7.80 &  1.26 &  1.10 
    \end{tabular}
    \caption{Yields for \PgUb mesons in \PbPb collisions in $\pt < 30\GeVc$ and $\abs{y} < 2.4$, corrected for acceptance and efficiency, and normalized by the nuclear thickness function $\langle \TAA \rangle$ and the number of minimum bias events $\NMB$. The values for the yields and their uncertainties are in units of pb$/\GeVc$.}
    \label{table:yield_cent_2S}
  \end{table}

\begin{table}[!hp]
\centering
    \label{table:yield_pt_3S}
    \begin{tabular}{cccc}
      \pt range (\GeVc) & $\frac{1}{\langle \TAA \rangle}\frac{\rd^2 N^{\PgU}}{\rd\pt\rd y}$ & Stat. Unc.               & Syst. Unc.               \\
      \hline
      0--4 &   1.11 & 0.44 & 0.24 \\
      4--9 &   1.04 & 0.39 & 0.24 \\
      9--15 &   0.38 & 0.14 & 0.11 \\
      15--30 &   0.07 & 0.02 & 0.01 
    \end{tabular}
    \caption{Yields for \PgUc mesons in \PbPb collisions in centrality 0--90\% and $\abs{y} < 2.4$, corrected for acceptance and efficiency, and normalized by the nuclear thickness function $\langle \TAA \rangle$ and the number of minimum bias events $\NMB$. The values for the yields and their uncertainties are in units of pb$/\GeVc$.}
\end{table}

\begin{table}[!hp]
\centering
    \label{table:yield_cent_3S}
    \begin{tabular}{ccccc}
      Centrality & $\langle \Npart \rangle$ & $\frac{1}{\langle \TAA \rangle}\frac{\rd^2 N^{\PgU}}{\rd\pt\rd y}$ & Stat. Unc. & Syst. Unc.               \\
      \hline
      0--30\% &  269.10    &  0.33 & 0.10 & 0.07  \\
      30--50\% &  109.10   &  0.70 & 0.19 & 0.11  \\
      50--70\% &  42.81  &  1.85 & 0.36 & 0.21  \\
      70--90\% &  11.43  &  3.42 & 1.05 & 0.68 
    \end{tabular}
    \caption{Yields for \PgUc mesons in \PbPb collisions in $\pt < 30\GeVc$ and $\abs{y} < 2.4$, corrected for acceptance and efficiency, and normalized by the nuclear thickness function $\langle \TAA \rangle$ and the number of minimum bias events $\NMB$. The values for the yields and their uncertainties are in units of pb$/\GeVc$.}
\end{table}

\begin{table}[!hp]
  \centering
    \begin{tabular}{cccc}
      Centrality & $\frac{1}{\langle \TAA \rangle}\frac{\rd^2 N^{\PgU}}{\rd\pt\rd y}$ & Stat. Unc. & Syst. Unc.               \\
      \hline
      0--5\%      & 15.20 & 0.89 & $\begin{matrix} +0.88 \\ -0.95 \end{matrix}$  \\
      5--10\%     & 15.28 & 1.01 & $\begin{matrix} +0.89 \\ -0.96 \end{matrix}$  \\
      10--20\%    & 15.40 & 0.82 & $\begin{matrix} +0.94 \\ -1.00 \end{matrix}$  \\
      20--30\%    & 19.12 & 1.86 & $\begin{matrix} +1.22 \\ -1.27 \end{matrix}$  \\
      30--40\%    & 23.06 & 1.29 & $\begin{matrix} +1.68 \\ -1.71 \end{matrix}$  \\
      40--50\%    &  24.95 & 1.70 & $\begin{matrix} +2.18 \\ -2.14 \end{matrix}$  \\
      50--60\%    &  28.96 & 2.51 & $\begin{matrix} +3.10 \\ -3.00 \end{matrix}$  \\
      60--70\%    &  43.89 & 4.17 & $\begin{matrix} +6.04 \\ -5.56 \end{matrix}$  \\
      70--100\%   &  37.69 & 6.21 & $\begin{matrix} +6.68 \\ -4.61 \end{matrix}$
    \end{tabular}
    \caption{Yields for \PgUa mesons in \PbPb collisions in $\pt < 30\GeVc$ and $\abs{y} < 2.4$, corrected for acceptance and efficiency, and normalized by the nuclear thickness function $\langle \TAA \rangle$ and the number of minimum bias events $\NMB$ from Ref.~\cite{HIN16023}. The values for the yields and their uncertainties are in units of pb$/\GeVc$.}
    \label{table:yield_cent_1S}
  \end{table}

}\cleardoublepage \section{The CMS Collaboration \label{app:collab}}\begin{sloppypar}\hyphenpenalty=5000\widowpenalty=500\clubpenalty=5000
\cmsinstitute{Yerevan Physics Institute, Yerevan, Armenia}
{\tolerance=6000
A.~Tumasyan\cmsAuthorMark{1}\cmsorcid{0009-0000-0684-6742}
\par}
\cmsinstitute{Institut f\"{u}r Hochenergiephysik, Vienna, Austria}
{\tolerance=6000
W.~Adam\cmsorcid{0000-0001-9099-4341}, J.W.~Andrejkovic, T.~Bergauer\cmsorcid{0000-0002-5786-0293}, S.~Chatterjee\cmsorcid{0000-0003-2660-0349}, K.~Damanakis\cmsorcid{0000-0001-5389-2872}, M.~Dragicevic\cmsorcid{0000-0003-1967-6783}, A.~Escalante~Del~Valle\cmsorcid{0000-0002-9702-6359}, P.S.~Hussain\cmsorcid{0000-0002-4825-5278}, M.~Jeitler\cmsAuthorMark{2}\cmsorcid{0000-0002-5141-9560}, N.~Krammer\cmsorcid{0000-0002-0548-0985}, L.~Lechner\cmsorcid{0000-0002-3065-1141}, D.~Liko\cmsorcid{0000-0002-3380-473X}, I.~Mikulec\cmsorcid{0000-0003-0385-2746}, P.~Paulitsch, F.M.~Pitters, J.~Schieck\cmsAuthorMark{2}\cmsorcid{0000-0002-1058-8093}, R.~Sch\"{o}fbeck\cmsorcid{0000-0002-2332-8784}, D.~Schwarz\cmsorcid{0000-0002-3821-7331}, S.~Templ\cmsorcid{0000-0003-3137-5692}, W.~Waltenberger\cmsorcid{0000-0002-6215-7228}, C.-E.~Wulz\cmsAuthorMark{2}\cmsorcid{0000-0001-9226-5812}
% ...
\par}
\vskip\cmsinstskip
\dag:~Deceased\\
$^{1}$Also at Yerevan State University, Yerevan, Armenia\\
$^{2}$Also at TU Wien, Vienna, Austria\\
% ...
\end{sloppypar}
\end{document}